\documentclass[12pt]{iopart}
\usepackage{iopams}
\usepackage{epsfig,cite}
\newcommand{\be}{\begin{equation}}
\newcommand{\ee}{\end{equation}}
\newcommand{\bd}{\begin{displaymath}}
\newcommand{\ed}{\end{displaymath}}
\newcommand{\BE}{\begin{eqnarray}}
\newcommand{\EE}{\end{eqnarray}}

\newcommand{\bra}{\left\langle}
\newcommand{\ket}{\right\rangle}

\newcommand{\erf}{{\rm erf}}

\newcommand{\bu}{\ensuremath{\mathbf{u}}}

\newcommand{\bsy}[1]{\boldsymbol{#1}}
\newcommand{\avg}[1]{\left\langle{#1}\right\rangle}
\newcommand{\davg}[1]{\left\langle\!\left\langle{#1}\right\rangle\!\right\rangle}
\newcommand{\ii}{{\rm i}}
\newcommand{\ovl}[1]{\overline{#1}}
\renewcommand{\l}{\left}
\renewcommand{\r}{\right}
\newcommand{\erfc}{{\rm erfc}}

\newcommand{\boldpsi}{{\mbox{\boldmath $\psi$}}}

\newcommand{\tp}{{t^\prime}}

\newcommand{\Rbar}{\overline{R}}
\newcommand{\Rtilde}{\widetilde{R}}

\begin{document}

\title[Dynamics of adaptive agents with asymmetric
information]{Dynamics of adaptive agents with asymmetric information}

\author{Andrea De Martino\dag, Tobias Galla\ddag
}

\address{\dag\ INFM SMC and Dipartimento di Fisica,
  Universit\`a di Roma ``La Sapienza'', p.le A. Moro 2, 00185 Roma (Italy)\\
  \ddag\ The Abdus Salam International Centre for Theoretical Physics,
  Strada Costiera 11, 34014 Trieste (Italy) and INFM, Trieste-SISSA
  Unit, v. Beirut 2-4, 34014 Trieste (Italy)}

\begin{abstract}
  We apply path-integral techniques to study the dynamics of
  agent-based models with asymmetric information structures. In
  particular, we devise a batch version of a model proposed originally
  by Berg et al. \cite{Bergetal}, and convert the coupled multi-agent
  processes into an effective-agent problem from which the dynamical
  order parameters in ergodic regimes can be derived self-consistently
  together with the corresponding phase structure. Our dynamical study
  complements and extends the available static theory. Results are
  confirmed by numerical simulations.
\end{abstract}


\ead{\tt Andrea.DeMartino@roma1.infn.it, galla@ictp.trieste.it}

\section{Introduction}

Over the past years the study of agent-based models of financial markets or other phenomena with the tools of statistical mechanics has proved to be
exceptionally fruitful. Many such models can, in
the language of statistical mechanics, be understood as fully
connected mean-field systems comprising disordered interactions and
various types of global frustration. They are thus perfectly suited to
be addressed with the techniques developed originally for other
purposes, such as the study of magnetic systems, spin-glasses or
neural networks. Indeed, both static and dynamical methods, including
replica techniques and path integrals, have been successfully applied
for example to the Minority Game (MG)
\cite{ChalZhan97}, presumably one of the most studied models in this
context, and have led to an advanced theoretical understanding of the
behaviour of various versions of the MG \cite{Book1,Book2,Book3}. On
the other hand these studies have also revealed new types of
complexity and phase transitions, which hitherto had been unknown to
statistical physicists.

The standard versions of the MG describe an ensemble of interacting
agents who at each time step react to publicly available information
by taking trading decisions, e.g. to buy or to sell a given asset.
While it is crucial in this setting that the information made
available is identical for all agents, no interaction
between the individual traders other than through this global and
uniform signal is present in the MG.  Furthermore, it is essentially
of little relevance whether this stream of information is generated
endogenously by the system or drawn externally at random \cite{Cava99}
as long as all agents react to the same signal. This suffices to
generate a remarkably complex dynamics with phenomena including phase
transitions, non-ergodic regimes, replica symmetry breaking and memory
effects .

The aim of the present paper is to extend the dynamical path-integral
formalism to the study of models with private, agent-dependent
information. These cases play a major role in economic theory,
especially in view of the connection between asymmetric information
and the failure of market equilibrium \cite{asymminfo}. The model we
address here was first introduced in \cite{Bergetal} and is a close
relative of the Shapley-Shubik model of non-cooperative trading
equilibrium \cite{ShapShub}. While the focus of \cite{Bergetal} lies
mainly on the statics of the model using replica techniques, we will
here complement this work by an analysis of the dynamics based on
generating functionals for systems with quenched disorder
\cite{DeDom78}. Although the dynamical analysis presented here
parallels that of the MG, the model displays some novel features and
new types of phases. Our analysis sets the stage for further studies
addressing subtle issues related to fluctuations and dependence on the
learning rate, inherently dynamical features which statics is unable
to capture. In the present paper we will mostly be concerned with the
mathematical analysis of the model, details of the economic background
can be found in \cite{Bergetal} and references therein.

\section{Model definitions}

The definition of the model follows closely that given in
\cite{Bergetal}. One considers a single-asset market with $N$ agents
labelled by Roman indices. It is assumed that the asset pays a
monetary return $R(\ell)$ at the end of each market round
$\ell=0,1,2,\dots$. This return depends on the value of a discrete
variable $\omega(\ell)$, which models the ``state of the world'' and is
similar to the global signal made available to agents in MGs. We will
here assume that $\omega(\ell)$ is determined externally, similar to MGs
with randomly drawn exogenous information.  Specifically, $\omega(\ell)$
is selected randomly and independently at each round with flat
probability distribution from the set $\{1,\dots,\Omega\}$. The
statistical mechanics analysis will ultimately be concerned with the
thermodynamic limit $N\to\infty$, which is taken in a way such that
the relative number of possible states of the world $\Omega/N$ remains
finite.  The ratio $\alpha=\Omega/N$ turns out to be the key control
parameter of the model.  The asset return at time $\ell$ is then given as
$R(\ell)=R^{\omega(\ell)}$, where the components of the vector
$\bsy{R}=(R^1,\dots,R^\Omega)$ are taken to be quenched random
variables, drawn at the start of the game and then kept fixed.  It is
assumed that they each take the form
\begin{equation}\label{err}
R^\omega=\Rbar+\frac{\Rtilde^\omega}{\sqrt{N}},
\end{equation}
where $\Rbar>0$ is a constant, and where the $\{\Rtilde^\omega\}$ are
independent, identically distributed Gaussian random variables with
zero mean and variance $\lambda^2>0$. Thus, the Bernoulli process
$\{\omega(\ell)\}_{\ell\geq 0}$ induces the time series of asset returns
$\{R(\ell)\equiv R^{\omega(\ell)}\}_{\ell\geq 0}$.  $\Rbar$ and $\lambda$ are
additional model parameters. The motivation for the choice (\ref{err})
will become clear in the following.

Contrary to MGs, traders in this model are unable to observe the state
$\omega(\ell)$ directly. Rather, each of them has access only to a
coarse-grained signal on $\{1,\ldots,\Omega\}$ which corresponds to
some fixed private information scheme. In particular, the signal
observed by a given trader $i$ is determined by the vector
\begin{equation}\label{info}
\bsy{k}_i:\{1,\ldots,\Omega\}\ni\omega\to k_i^\omega\in\{-1,1\}.
\end{equation}
The components of any $\bsy{k}_i$ ($i=1,\ldots,N$) are
again assumed to be drawn at random and independently from $\{-1,1\}$
with equal probability for all $i$ and $\omega$ at the beginning of
the game and are kept fixed afterwards. In this way, each trader
has a private information source providing him with a binary signal
$k_i^{\omega(\ell)}$ at time $\ell$. This private signal does depend on the
state $\omega(\ell)$, but exactly what this state is at time $\ell$ is not
known to the individual agents. Since the $\{R^\omega\}$ and the
$\{\bsy{k}_i\}$ are drawn independently, the sequence
$\{k_i^{\omega(\ell)}\}_{\ell\geq 0}$ observed by agent $i$ will in general
not allow him to tell what return $R^{\omega(\ell)}$ is to be expected at
time step $\ell$. Crucially, however, the correlation between the vectors
$\bsy{k}_i$ and $\bsy{R}$ are heterogeneous across the population of
traders, so that different agents will have varying
abilities to resolve the individual states
$\omega\in\{1,\dots,\Omega\}$.

At each round $\ell$ of the game every agent decides to invest a
monetary amount $z_{i k_i^{\omega(\ell)}}(\ell)\geq 0$ which depends
on the signal $k_i^{\omega(\ell)}\in\{-1,1\}$ he receives at stage
$\ell$. The total amount invested by agents at round $\ell$ determines
the price of the asset: \be\label{eq:price}
p(\ell)=\frac{1}{N}\sum_{i=1}^N\sum_{\sigma\in\{-1,1\}}
z_{i\sigma}(\ell)\delta_{\sigma,k_i^{\omega(\ell)}}. \ee It remains to
specify how the agents determine the amounts $\{z_{i\sigma}(\ell)\}$
they invest.  It is assumed that traders are adaptive and that their
behaviour is governed by an inductive learning dynamics. Specifically,
each agent has a propensity $u_{i\sigma}(\ell)$ to invest under each
of the two possible signals $\sigma\in\{-1,1\}$; these propensities
are initialized at values $u_{i\sigma}(0)$ at time $t=0$ and are
updated at the end of every round according to the marginal success of
the investment: \be
u_{i\sigma}(\ell+1)=u_{i\sigma}(\ell)+\Gamma\frac{\partial
Q_i(\ell)}{\partial z_{i\sigma}(\ell)}.  \ee Here $\Gamma>0$ is a
learning rate\footnote{In principle, different agents $i$ might have
different learning rates $\Gamma_i$. We here restrict the discussion
to the case $\Gamma_i\equiv\Gamma$ for all $i$. Our theory can be
generalised to populations of agents with heterogeneous learning
rates.}, while $Q_i(\ell)$ stands for the payoff received by trader
$i$ at the end of round $\ell$. The $\{Q_i(\ell)\}$ can be specified
upon noting that at each round every agent puts forward his bid before
$p(\ell)$ is known and makes profit if the return of the asset exceeds
its price.  The price in turn is determined by the collective action
of all traders, Eq.  (\ref{eq:price}). Now since agent $i$ acquires
$z_{i,k_i^{\omega(\ell)}}(\ell)/p(\ell)$ units of the asset, each
yielding an effective return of $R(\ell)-p(\ell)$, the following form
for the payoffs is assumed:\BE\label{eq:payoff} Q_i(\ell)=
\frac{z_{i\sigma}(\ell)}{p(\ell)} \left[R(\ell)-
p(\ell)\right]\delta_{\sigma,k_i^{\omega(\ell)}}. \EE For simplicity,
we linearise the payoff as in \cite{Bergetal} and consider \be
Q_i(\ell)=z_{i\sigma}(\ell) \left[R(\ell)-
p(\ell)\right]\delta_{\sigma,k_i^{\omega(\ell)}}. \ee To compute the
marginal payoffs $\frac{\partial Q_i(\ell)}{\partial
z_{i\sigma}(\ell)}$, note that $p(\ell)$ is itself a function of
$z_{i\sigma}(\ell)$, so that in principle one has \be \frac{\partial
Q_i(\ell)}{\partial z_{i\sigma}(\ell)}= \left[R(\ell)-
p(\ell)\right]\delta_{\sigma,k_i^{\omega(\ell)}}-z_{i\sigma}(\ell)\frac{
\partial p(\ell)}{\partial
z_{i\sigma}(\ell)}\delta_{\sigma,k_i^{\omega(\ell)}}.  \ee In general,
traders may not be able to evaluate how much their decision affects
the price, i.e. na\"\i ve agents who neglect the impact of their own
trading action might ignore the second term on the right-hand side,
while other so-called sophisticated agents \cite{MarsChalZecc00} might
be able to take into account this market-impact correction. We will
not allow for any heterogeneity across the agents in this respect, but
will instead study the dynamics \be\label{eq:onlineupdate}
u_{i\sigma}(\ell+1)=u_{i\sigma}(\ell)+\Gamma\delta_{\sigma,k_i^{\omega(\ell)}}
\left[R(\ell)-p(\ell)-\frac{\eta}{N}z_{i\sigma}(\ell)\right]. \ee Here
$\eta\geq 0$ is a further model parameter which accounts for the
(uniform) ability of the agents to estimate the impact of their
trading actions on the price. For $\eta=0$ agents act as `price
takers' and neglect the effect of the term $\frac{ \partial
p(\ell)}{\partial z_{i\sigma}(\ell)}$ when updating the propensities;
on the other hand for $\eta=1$ they fully correct for their impact on
the price process. Tuning $\eta\in[0,1]$ allows to consider agents of
increasing sophistication, similar to what is done in MGs with
market-impact correction. Finally, the propensities
$\{u_{i\sigma}(\ell)\}$ are related to the investments
$\{z_{i\sigma}(\ell)\}$ made at time step $\ell$ via \be\label{mapp}
z_{i\sigma}(\ell)=u_{i\sigma}(\ell)\theta[u_{i\sigma}(\ell)], \ee
where $\theta(x)$ is the step-function, i.e. $\theta(x)=1$ for $x>0$
and $\theta(x)=0$ otherwise\footnote{In \cite{Bergetal} it is argued
that the key features of the model remain unchanged for other choices
for the map $u_{i\sigma}(\ell)=u[z_{i\sigma}(\ell)]$, as long as this
map is increasing and guarantees $\lim_{u\to\infty}z(u)=\infty$ and
$\lim_{u\to-\infty}z(u)=0$. In the simulations presented below the
expression given in (\ref{mapp}) was used. }.

The model may thus be summarised as follows. At each time step $\ell$ a
`state of the world' $\omega(\ell)\in\{1,\ldots,\Omega\}$ is drawn at
random. Each agent $i$ then receives a signal
$k_i^{\omega(\ell)}\in\{-1,1\}$ corresponding to his private information
structure. Given the perceived signal he decides to invest an amount
$z_{i,k_i^{\omega(\ell)}}(\ell)\geq 0$ according to (\ref{mapp}), based on the propensity $u_{i,k_i^{\omega(\ell)}}(\ell)$.  From the
investments of all agents a total price $p(\ell)$ is determined via the
market clearing condition (\ref{eq:price}).  Each individual agent $i$
then updates the propensity $u_{i\sigma}$ for the
perceived signal $\sigma=k_i^{\omega(\ell)}$ according to
(\ref{eq:onlineupdate}), and leaves the propensity for the opposite signal $-\sigma$ unchanged.

We shall mostly be interested in the stationary state(s) of the model, and in particular in the question of whether agents can
co-ordinate efficiently even if their access to the global signal is
filtered through their private information schemes. We will refer
to `market efficiency' as the situation in which returns are fully
reflected by the prices. In order to
quantify the efficiency, or otherwise, of the market we consider the
`squared distance' between prices and returns in the steady
state\footnote{It will turn out that
  $\avg{R(\ell)-p(\ell)|\omega}=\Or(N^{-1/2})$, so that $H$ as defined above
  is indeed of order one. This is observed in numerical simulations,
  but will also become manifest in the course of the generating
  functional analysis presented below.}: \be
H=\lim_{N\to\infty}\sum_\omega\avg{R(\ell)-p(\ell)|\omega}^2, \ee where
$\avg{\cdots|\omega}$ stands for a time-average in the steady state
conditioned on the occurrence of state $\omega$.  Phases in which
prices equal returns in all states $\omega$, and in which accordingly
$H=0$ are said to be efficient, whereas phases with $H\neq 0$ are
referred to as inefficient.

A second quantity of interest measures how differently the agents
behave upon receiving the two different signals, and is defined as
follows: \be\label{eq:qdef}
q=\frac{1}{N}\sum_{i=1}^N\left(\frac{\avg{z_{i+}}-\avg{z_{i-}}}{2}\right)^2.
\ee Here $\avg{\dots}$ stands for a time-average in the stationary
state. $q$ measures the extent to which agents use the information
available to them. Small values of $q$ indicate that the investment
they make is nearly independent of the observed signal, while the
perceived signal is strongly correlated with their trading actions for
large values of $q$.

Before turning to the details of the further analysis, let us briefly
summarise the static picture of the model as found in
\cite{Bergetal}. For $\eta=0$ the system undergoes a phase transition
from an efficient to an inefficient regime at a critical value
$\alpha_c$, which depends on the details of the disorder
statistics. In the sub-critical phase at $\alpha<\alpha_c$, where
$H=0$, the asymptotic value of $q$ depends on initial conditions. For
the game with sophisticated agents ($\eta=1$), instead, no such
transition occurs and the system is inefficient for all
$\alpha$. Moreover the dynamics is ergodic. Loosely writing $q$ turns
out to be larger for the game without impact-correction than it is for
$\eta=1$. Thus, na\"\i ve agents use the information more than
sophisticated ones, though the latter have larger gains.

\section{Generating functional analysis}
\subsection{Batch dynamics}

\begin{figure}[t]
\vspace*{1mm}
\begin{tabular}{cc}
\epsfxsize=70mm  \epsffile{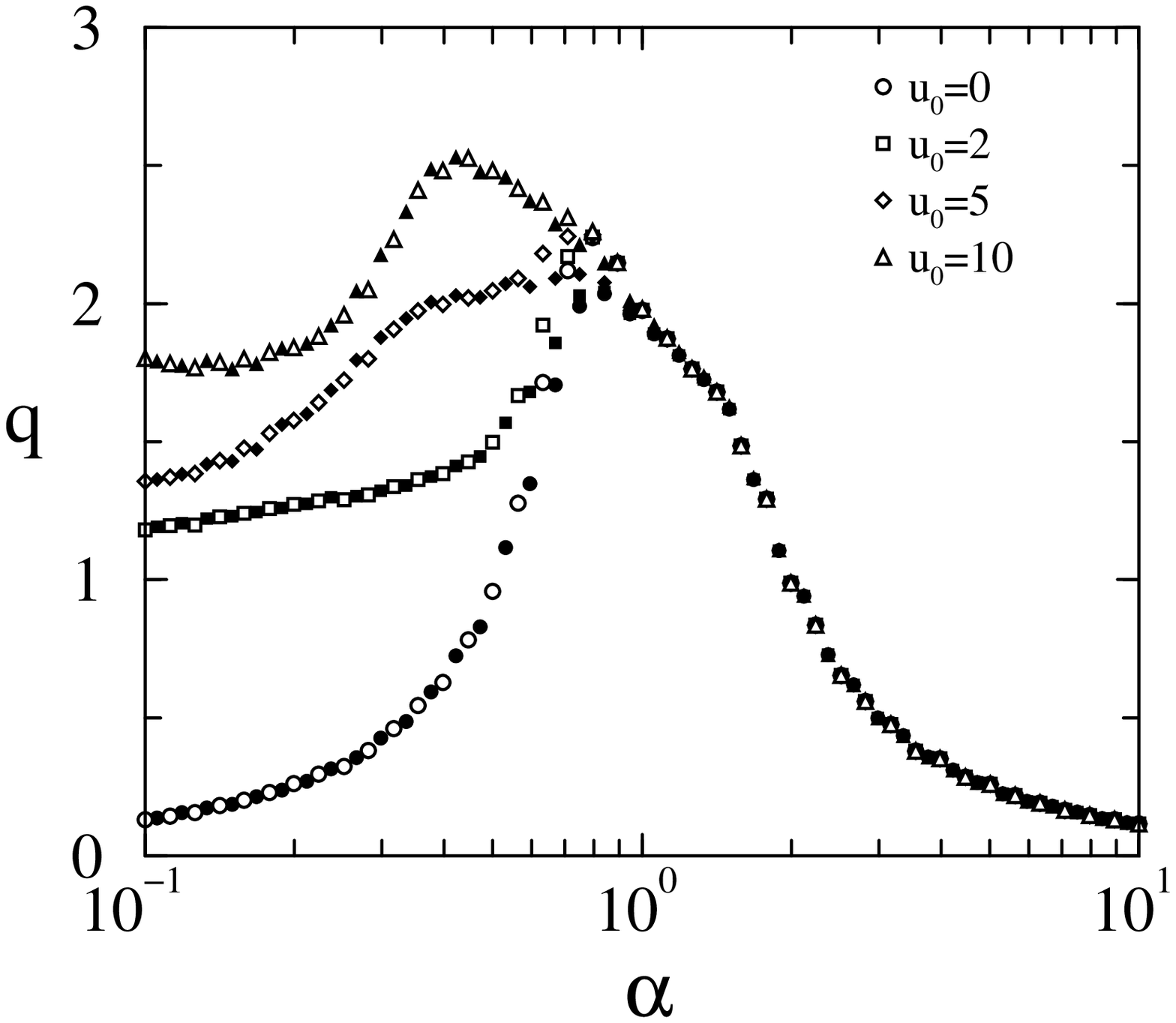} ~~&~~
\epsfxsize=70mm  \epsffile{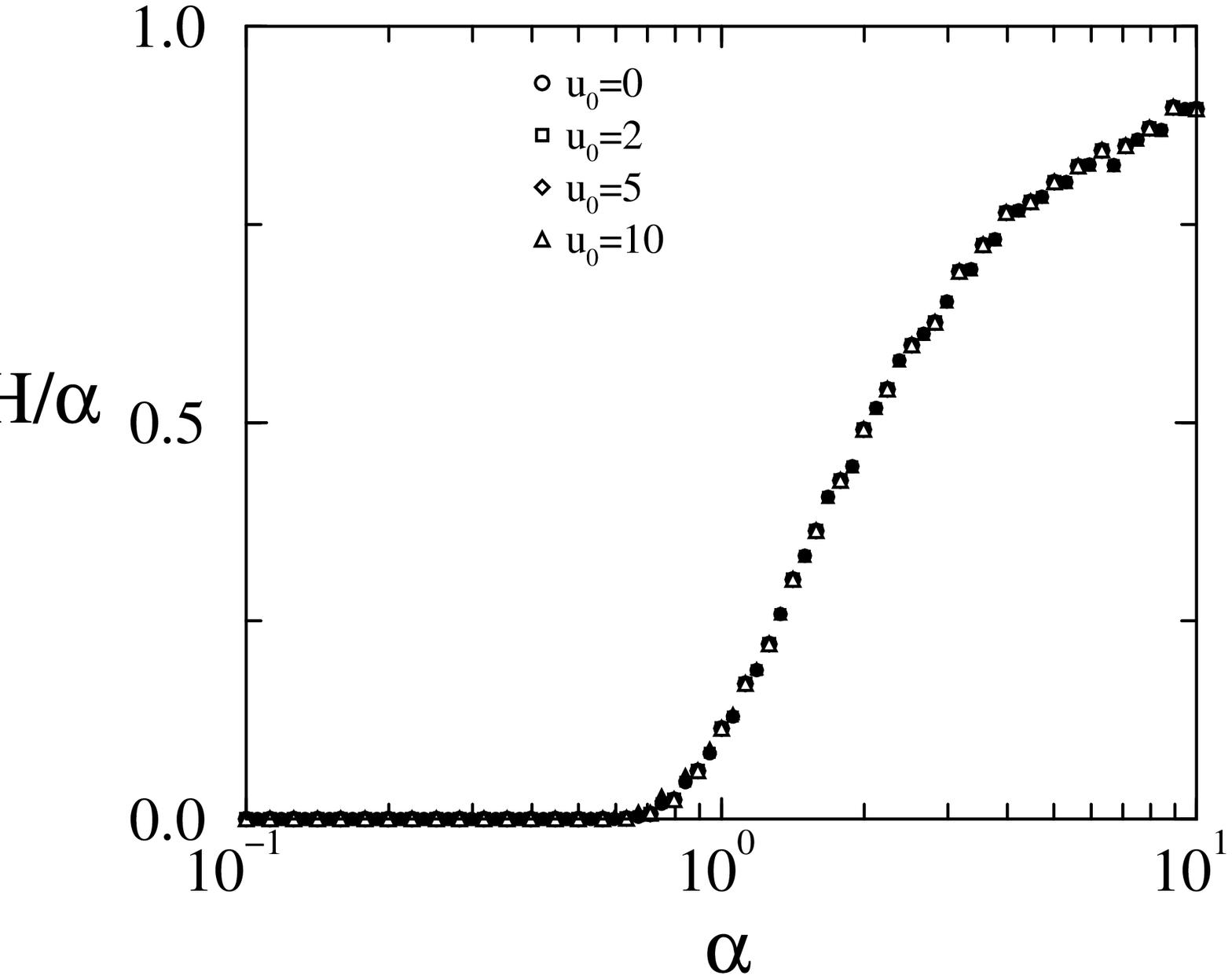}
\end{tabular}
\vspace*{4mm} \caption{Order parameters $q$ and $H/\alpha$ vs.
  $\alpha$ at fixed $\Rbar=\Gamma=\lambda=1$ for the batch and on-line games
  ($\eta=0$), for different initial conditions, $u_{i\sigma}(0)=\sigma
  u_0$ for all $i$. Open symbols are results from simulations of the
  on-line game, solid symbols correspond to the batch process.
  Numerical simulations are performed at $\alpha N^2=10^4$ and all
  data are averages over $100$ samples of the disorder. Measurements
  are taken over $5\cdot 10^4$ steps in the on-line and $7500$ steps
  in the batch games, respectively, preceded by $15\cdot 10^4$
  equilibration steps in the on-line case, and $22500$ equilibration
  steps for the batch dynamics. }\label{fig:batcheqonline}
\end{figure}

The learning rule (\ref{eq:onlineupdate}) corresponds to what is known
as an `on-line' model in the context of the MG \cite{Book2}, with an
explicit dependence on $\omega(\ell)$ at each time step $\ell$. It is
analytically convenient to replace this type of updating scheme by one
in which an effective average over all values of
$\omega(\ell)\in\{1,\dots,\alpha N\}$ is carried out at every time
step, as in MGs \cite{Book2}. The resulting `batch' model roughly
describes a situation in which propensity updates are performed only
once every $\Or(\Omega)$ time steps, and is defined by
\be\label{eq:batchupdatenotrecaled}
u_{i\sigma}(\ell+1)=u_{i\sigma}(\ell)+\frac{\Gamma}{\Omega}\sum_{\omega=1}^{\Omega}
\delta_{\sigma,k_i^\omega}\bigg[R^\omega-p^\omega(\ell)-\frac{\eta}{N}
z_{i\sigma}(\ell)\bigg], \ee where now the price at round $\ell$ is
effectively a vector of prices, $\{p^\omega(\ell)\}$, one for each
state $\omega$: $N p^\omega(\ell)=\sum_i z_{i,k_i^\omega}(\ell)$. This
modification has turned out to have little effect on the steady state
properties of conventional MGs \cite{HeimCool01}, but carries the
advantage of simplifying the analytical theory considerably. Numerical
results confirm that the batch and on-line versions of the present
model are qualitatively and quantitatively very similar as far as the
order parameters $q$ and $H$ and the breakdown of ergodicity at
$\alpha_c$ are concerned, see Fig. \ref{fig:batcheqonline}.
Notice that in the present batch model the learning rate $\Gamma$
effectively fixes a time-scale, it also has a subtle influence on
transients. It is convenient to introduce the variables
\be\label{defs} x_i(\ell)=\frac{1}{2}\sum_{\sigma\in\{-1,1\}}
z_{i\sigma}(\ell),~~~~~~~~~~ y_i(\ell)=\frac{1}{2}\sum_{\sigma\in\{-1,1\}}
\sigma z_{i\sigma}(\ell), \ee in terms of which one has
$z_{i\sigma}(\ell)=x_i(\ell)+\sigma y_i(\ell)$, and
\BE
\fl u_{i\sigma}(\ell+1)=u_{i\sigma}(\ell)+\frac{\Gamma}{\Omega}\sum_\omega
\delta_{\sigma,k_i^\omega}\l[\Rbar-\frac{1}{N}\sum_j x_j(\ell)\r]
-\frac{\Gamma\eta}{\Omega N}\sum_\omega
\delta_{\sigma,k_i^\omega} z_{i\sigma}(\ell)\nonumber\\
+\frac{\Gamma}{\Omega\sqrt{N}}\sum_\omega
\delta_{\sigma,k_i^\omega}\l[\Rtilde^\omega-\frac{1}{\sqrt{N}}\sum_j k_j^\omega y_j(\ell)\r]
+h_{i\sigma}(\ell),
\label{eq:batchupdate}
\EE where we have introduced external perturbation fields
$\{h_{i\sigma}(\ell)\}$ which will be used later in order to generate
response functions. While the dynamics of the MG with $S=2$ strategies
per player can be written in terms of one degree of freedom per agent
(the so-called score difference), no such simplification can be made
in the present game and we have to keep both the $\{x_i(\ell)\}$ and
the $\{y_i(\ell)\}$. Writing $\delta_{\tau\tau'}=(1+\tau\tau')/2$ for
$\tau,\tau\in\{-1,1\}$, (\ref{eq:batchupdate}) can be simplified
further using the facts that $\lim_{\Omega\to\infty}
(1/\Omega)\sum_\omega\Rtilde^\omega=0$ and
$\lim_{\Omega\to\infty}(1/\Omega)\sum_\omega
\delta_{\sigma,k_i^\omega}=1/2$. Subsequently, one may re-scale time
to arrive at \BE\label{by} \fl
u_{i\sigma}(t+1)=u_{i\sigma}(t)+\frac{\Gamma}{N}\sum_\omega
\delta_{\sigma,k_i^\omega}\sum_j\l[\Rbar- x_j(t)\r]+
\frac{\Gamma\sigma}{2\sqrt{N}}\sum_\omega k_i^\omega \Rtilde^\omega
-\frac{1}{2}\Gamma\eta\alpha z_{i\sigma}(t)\nonumber\\
-\frac{\Gamma}{N}\sum_\omega \delta_{\sigma,k_i^\omega}\sum_j
k_j^\omega y_j(t)+h_{i\sigma}(t),\EE where $t$ is now a re-scaled
time. (\ref{by}) is the process we shall consider.

\subsection{The effective-agent process}

The aim of every dynamical theory of disordered systems is to derive a
closed set of equations for the behavior of macroscopic correlation
and response functions in the limit $N\to\infty$. We apply the
path-integral method first devised for systems with quenched disorder
by De Dominicis \cite{DeDom78}. This technique is based on the
computation of the moment generating functional \BE Z[\boldpsi]
&=\davg{\exp\l[\ii\sum_{i,\sigma,t}
\psi_{i\sigma}(t)z_{i\sigma}(t)\r]}\nonumber \\ & =\int D\bsy{u}
~p(\bsy{u}(0)) ~\exp\l[\ii\sum_{i,\sigma,t}
\psi_{i\sigma}(t)z_{i\sigma}(t)\r]\prod_{t,\sigma,i}\delta\l[{\rm
equation~(\ref{by})}\r],\label{eq:genfct} \EE where
$\boldpsi=(\boldpsi_-,\boldpsi_+)$ are generating source fields and
where $\davg{\cdots}$ denotes an average over the process
(\ref{eq:batchupdate}) with respect to the distribution
$p(\bsy{u}(0))$ of initial conditions $\bsy{u}(0)=\{u_{i\sigma}(0)\}$
from which the dynamics is started. The shorthand $D\bsy{u}$ stands
for $\prod_{t,\sigma,i}
\left[du_{i\sigma}(t)/\sqrt{2\pi}\right]$. Correlation and response
functions can be obtained by taking derivatives of the
disorder-averaged generating functional $\ovl{Z[\bsy{\psi}]}$ with
respect to the source fields $\{\psi_{i\sigma}(t)\}$ and/or the
perturbing fields $\{h_{i\sigma}(t)\}$ and by subsequently considering
the limit of vanishing fields. Our ultimate goal is to obtain
self-consistent equations for these physically relevant observables in
the limit $N\to\infty$.

The calculation can be performed along the following lines. First, one
expresses the $\delta$-distributions in (\ref{eq:genfct}) in their
exponential representation (we denote the conjugate parameters of the
$\{u_{i\sigma}(t)\}$ by $\{\widehat{u}_{i\sigma}(t)\}$). Next, the
average over the quenched disorder $\{\bsy{R}\}$ and $\{\bsy{k}_i\}$
is evaluated. This is done conveniently by introducing the parameters
\be\label{bel} a^\omega(t)=\frac{1}{\sqrt{N}}\sum_i y_i(t)
k_i^\omega,~~~~~~~~ b^\omega(t)=\frac{1}{\sqrt{N}}\sum_i w_i(t)
k_i^\omega, \ee where $y_i(t)$ is given by (\ref{defs}) while $w_i(t)$
stands for \be w_i(t)=\frac{1}{2}\sum_{\sigma\in\{-1,1\}} \sigma
\widehat{u}_{i\sigma}(t). \ee As usual, disorder-averaging generates
macroscopic objects (dynamical order parameters). In the present
problem, they are given by \BE
m(t)=\frac{1}{\sqrt{N}}\sum_i\l[\Rbar-x_i(t)\r],~~~~
\varphi(t)=\frac{1}{2\sqrt{N}}\sum_{i,\sigma}\widehat{u}_{i\sigma}(t),
\nonumber\\Q(t,t')=\frac{1}{N}\sum_i y_i(t)y_i(t'),~~~~~~
L(t,t')=\frac{1}{N}\sum_i w_i(t)w_i(t'),\\K(t,t')=\frac{1}{N}\sum_i
y_i(t)w_i(t').\nonumber \EE These definitions can again be inserted
into $\ovl{Z[\bsy{\psi}]}$ via $\delta$-distributions. Finally,
factorising the resulting expression over agents and states wherever
possible one arrives at the familiar form \be \fl
\overline{Z[\boldpsi]}=\int
e^{N\l(\Psi+\Phi+\Upsilon\r)+\Or(\sqrt{N})} D\bsy{Q}
D\bsy{\widehat{Q}} D\bsy{K} D\bsy{\widehat{K}} D\bsy{L}
D\bsy{\widehat{L}} D\bsy{m} D\bsy{\widehat{m}} D\bsy{\varphi}
D\bsy{\widehat{\varphi}}. \label{gf}\ee The functions
$\Psi\equiv\Psi(\bsy{m},\bsy{\varphi},\bsy{Q},\bsy{\widehat{Q}},\bsy{L},\bsy{\widehat{L}},
\bsy{K},\bsy{\widehat{K}})$,
$\Phi\equiv\Phi(\bsy{m},\bsy{\varphi},\bsy{Q},\bsy{L},\bsy{K})$ and
$\Upsilon\equiv\Upsilon(\bsy{\widehat{m}},
\bsy{\widehat{\varphi}},\bsy{\widehat{Q}},\bsy{\widehat{L}},\bsy{\widehat{K}})$
are given respectively by \BE \fl
\Psi=\ii\sum_{t,t'}\left[Q(t,t')\widehat{Q}(t,t')+
L(t,t')\widehat{L}(t,t')+K(t,t')\widehat{K}(t,t')\right]-\ii\Gamma\alpha\sum_tm(t)\varphi(t),\\
\fl \Phi=\alpha\log \int Da D\widehat{a}Db D\widehat{b}~
\exp\l[S(a,\widehat{a},b,\widehat{b})\r],\label{mops}\\ \fl
\Upsilon=\frac{1}{N}\sum_i\log \int Du D\widehat{u}~
\left[\prod_\sigma p_\sigma(u_\sigma(0))\right]
\exp\l[\ii\sum_{t,\sigma}\psi_{i\sigma}(t)z_\sigma(t)-\ii\sum_t
\widehat{m}(t)\l[\Rbar-x(t)\r]\r]\nonumber \\
\times\exp\left[-\ii\sum_{t,t'}\l[\widehat{Q}(t,t')y(t)y(t')
+\widehat{L}(t,t')w(t)w(t')+\widehat{K}(t,t')y(t)w(t')\r]\right]\nonumber\\
\times \exp\left[\ii\sum_{t,\sigma}\widehat{u}_\sigma(t)\left[
u_\sigma(t+1)-u_\sigma(t)-h_{i\sigma}(t)+\frac{1}{2}\widehat{\varphi}(t)+
\frac{1}{2}\Gamma\eta\alpha z_{\sigma}(t)\right]\right],
\label{eq:Omega}
\EE with the shorthands $Da=\prod_t \left[da(t)/\sqrt{2\pi}\right]$ (and similarly
for $D\widehat{a}$, $Db$, $D\widehat{b}$) and
$Du=\left[\prod_{t,\sigma}du_\sigma(t)/\sqrt{2\pi}\right]$ (and
similarly for $D\widehat{u}$), and \BE \fl
S(a,\widehat{a},b,\widehat{b})=\ii\sum_t\l[a(t)\widehat{a}(t)+b(t)\widehat{b}(t)+
\Gamma a(t)[b(t)+\varphi(t)]-\Gamma b(t)m(t)\r]\nonumber\\
-\frac{1}{2}\sum_{t,t'}\l[Q(t,t')\widehat{a}(t)\widehat{a}(t')+
L(t,t')\widehat{b}(t)\widehat{b}(t')+
2K(t,t')\widehat{a}(t)\widehat{b}(t')\r]\nonumber\\
-\frac{\Gamma^2\lambda^2}{2}\sum_{t,t'}b(t)b(t').  \EE Note that we
have assumed that the distribution of initial conditions factorizes 
over agents and the index $\sigma$ and that the distribution of the starting points is identical for all agents, so that $p(\bu(0))=\prod_{i\sigma}
p_\sigma(u_{i\sigma}(0))$. The integrals in (\ref{gf}) can
then be performed by the method of steepest descents in the limit
$N\to\infty$.  The dominant contributions here come from terms of
order $N$ in the exponent, and are contained in $\Psi$, $\Phi$ and
$\Upsilon$ as given above. All terms of sub-leading order carry zero
weight in the thermodynamic limit.  The saddle-point conditions can
then be worked out by taking derivatives of the exponent in (\ref{gf})
with respect to the integration variables. Differentiation with
respect to $\bsy{\widehat{m}}$ and $\bsy{\widehat{\varphi}}$ leads to
the relations
\BE\avg{x(t)-\Rbar}_\star=0~\Rightarrow~\frac{1}{2}\sum_\sigma
\avg{z_\sigma(t)}_\star=\Rbar,\label{pops}\\ \sum_\sigma \bra \hat
u_\sigma(t)\ket_\star=0, \EE where $\avg{\cdots}_\star$ denotes an
average with respect to the probability measure defined by the
single-agent process described by $\Upsilon$: \BE
\avg{\cdots}_\star=\frac{\int Du D\widehat{u}~
\left[\prod_\sigma p_\sigma(u_\sigma(0))\right] \cdots M(u,\widehat u)~e^{-\ii\sum_t
\widehat{m}(t)\l[\Rbar-x(t)\r]}}{ \int Du D\widehat{u}~
\left[\prod_\sigma p_\sigma(u_\sigma(0))\right]M(u,\widehat{u}) ~e^{-\ii\sum_t
\widehat{m}(t)\l[\Rbar-x(t)\r]}},\\ \fl M(u,\widehat{u})=
\exp\left[-\ii\sum_{t,t'}\l[\widehat{Q}(t,t')y(t)y(t')
+\widehat{L}(t,t')w(t)w(t')+\widehat{K}(t,t')y(t)w(t')\r]\right]\nonumber\\
\times \exp\left[\ii\sum_{t,\sigma}\widehat{u}_\sigma(t)\left[
u_\sigma(t+1)-u_\sigma(t)-h_\sigma(t)+\frac{1}{2}\widehat{\varphi}(t)+
\frac{1}{2}\Gamma\eta\alpha z_{\sigma}(t)\right]\right].\label{sap}
\EE All order parameters appearing in this measure take their
saddle-point values and the auxiliary fields $\bsy{\psi}$ have been
set to zero; also, we assumed $h_{i\sigma}(t)=h_\sigma(t)$ for all
$i$.  Notice that (\ref{pops}) implies that the average investment
equals $\Rbar$. We will henceforth set $\bsy{\widehat{m}}=0$ and use
(\ref{pops}) as an additional condition to be satisfied by the
solutions. As for $\bsy{\widehat{\varphi}}$, its physical meaning and
value will become clear when the steady state will be worked out
explicitly in the next section. The saddle-point conditions for
$\bsy{m}$ and $\bsy{\varphi}$ read \be m(t)=\avg{a(t)}_\Phi,~~~~~~~
\varphi(t)=-\avg{b(t)}_\Phi, \ee where \be
\avg{\cdots}_\Phi=\frac{\int \cdots
\exp\l[S(a,\widehat{a},b,\widehat{b})\r]
DaD\widehat{a}DbD\widehat{b}}{\int
\exp\l[S(a,\widehat{a},b,\widehat{b})\r]
DaD\widehat{a}DbD\widehat{b}}.  \ee These two equations admit the
self-consistent solution $\bsy{m}=\bsy{\varphi}=0$, the integrals
vanishing due to symmetry.

It remains to compute the saddle-point values of the order parameters
$\{\bsy{Q}, \bsy{L}, \bsy{K}\}$ and of their conjugates
$\{\bsy{\widehat{Q}}, \bsy{\widehat{K}},\bsy{\widehat{L}}\}$.
Extremisation of the exponent in (\ref{gf}) with respect to
$\{\bsy{\widehat{Q}}, \bsy{\widehat{K}},\bsy{\widehat{L}}\}$ gives \BE
Q(t,t')=\avg{y(t)y(t')}_\star=\frac{1}{4}\sum_{\sigma,\tau}
\sigma\tau\avg{z_\sigma(t) z_\tau(t')}_\star \label{qu},\\
L(t,t')=\avg{w(t)w(t')}_\star=\frac{1}{4}\sum_{\sigma,\tau}
\sigma\tau\avg{ \widehat{u}_\sigma(t) \widehat{u}_\tau(t')}_\star,\\
K(t,t')=\avg{y(t)w(t')}_\star=\frac{1}{4}\sum_{\sigma,\tau}
\sigma\tau\avg{ z_\sigma(t) \widehat{u}_\tau(t')}_\star.
\label{kapa}\EE By taking derivatives of the
generating functional with respect to the fields
$\{\bsy{\psi},\bsy{h}\}$ one may check that $\bsy{Q}$ and $\bsy{K}$
can be identified with correlation and response functions of the
original multi-agent system \BE Q(t,t')=
\lim_{N\to\infty}\frac{1}{N}\sum_{i=1}^N \ovl{\davg{y_i(t)
    y_i(t')}},\\\label{kaps}
K(t,t')=\ii\lim_{N\to\infty}\frac{1}{2N}\sum_{i=1}^N\sum_{\tau\in\{-1,1\}}
\tau~\frac{\partial\ovl{\davg{y_i(t)}}}{\partial h_{i\tau}(t')}.\EE
$\bsy{L}$ vanishes identically by virtue of the built-in normalization
$Z[\bsy{0}]=1$ (see \cite{HeimCool01}).  Finally, the saddle-point
equations corresponding to the integrations over
$\{\bsy{Q},\bsy{L},\bsy{K}\}$ read \be
\widehat{Q}(t,t')=\ii\frac{\partial\Phi}{\partial Q(t,t')},\quad
\widehat{L}(t,t')=\ii\frac{\partial\Phi}{\partial L(t,t')},\quad
\widehat{K}(t,t')=\ii\frac{\partial\Phi}{\partial K(t,t')}. \ee $\Phi$
can be evaluated explicitly by successively integrating (\ref{mops}) over the
$\{a(t)\}$, $\{\widehat{a}(t)\}$ and $\{b(t)\}$. Taking
the required derivatives one finds that \be
\bsy{\widehat{Q}}=\bsy{0},~~~~~~~
\bsy{\widehat{L}}=-\frac{\ii}{2}\alpha\bsy{\Lambda},~~~~~~~
\bsy{\widehat{K}}^T=-\alpha\Gamma\l(\bsy{1}-\ii\Gamma\bsy{K}\r)^{-1},
\ee where \BE \bsy{\Lambda}=\l(\bsy{1}-\ii\Gamma\bsy{K}\r)^{-1}
\bsy{D}\l(\bsy{1}-\ii\Gamma\bsy{K}^T\r)^{-1},\\
D(t,t')=\Gamma^2\l[\lambda^2+Q(t,t')\r]. \EE Motivated by (\ref{kaps}),
we shall henceforth set $\bsy{G}=-\ii\bsy{K}$.  Inserting the
saddle-point conditions into (\ref{sap}) leads to the effective-agent
process \BE\label{eq:effprocess} \fl
u_\sigma(t+1)=u_\sigma(t)+h_\sigma(t)-\frac{1}{2}\widehat{\varphi}(t)-\frac{\alpha\Gamma\sigma}{2}\sum_{\tp\leq
  t}(\bsy{1}+\Gamma
\bsy{G})^{-1}(t,t')y(t')\nonumber\\~~~~~~~~~~~~~~~~~~~~~~~
~~~~~~~~~~~~~~~~~~~~~~~~~~~~ -\frac{1}{2}\Gamma\eta\alpha
z_\sigma(t)+\frac{\sqrt{\alpha}}{2}\sigma\xi(t), \EE where
$\sigma\in\{-1,1\}$ and $z_\sigma(t)=u_\sigma(t)\theta[u_\sigma(t)]$,
while $\xi(t)$ is a zero-average Gaussian noise with temporal
correlations given by the covariance matrix \be\label{eq:correlator}
\avg{\xi(t)\xi(\tp)}=\Lambda(t,t').\ee Note that the
representative agent is here described by a pair of processes,
$u_\sigma(t)$, $\sigma\in\{-1,1\}$ at variance with the versions of the MG
studied so far, where one has only process (for the so-called score
difference).  The one-time function $\bsy{\widehat{\varphi}}$ and the
two-time functions $\bsy{Q}$ and $\bsy{G}$ are the dynamical order
parameters of the problem, to be determined self-consistently
according to (\ref{qu}) and (\ref{kapa}), where the average
$\avg{\cdots}_\star$ is over the effective process
(\ref{eq:effprocess}), i.e. over realisations of $\{\xi(t)\}$, subject
to the constraint (\ref{pops}). As usual, the resulting
self-consistent effective agent problem is fully equivalent to the
original coupled $N$-particle dynamics in the thermodynamic limit
$N\to\infty$ in the sense that $\bsy{Q}$ and $\bsy{G}$ are the
correlation and response functions of the original batch problem. The
non-trivial correlation structure of the single-particle noise and the
non-Markovian term coupling to times $\tp\leq t$ in the effective
process (\ref{eq:effprocess}) are direct consequences of the presence
of quenched disorder in the original multi-agent system, and impede a
full analytical solution for the dynamical order parameters at all
times. For models of the present type the alternative numerical
iteration scheme provided by \cite{EissOppe92} is restricted to
$\Or(10^2)$ time steps due to computational limitations. In our case
equilibration times turn out to be much larger. Note that the
constraint $\avg{x(t)}_\star=\Rbar$ (for all times $t$) is not an
external one as for example in spherical models
\cite{GallCoolSher03}, but rather it is self-generated by the system
in the thermodynamic limit.  For this reason $\bsy{\widehat{\varphi}}$
has no direct analogue in the original $N$-particle problem, but
appears only in the effective process.

\subsection{Average price/return fluctuations}

The fluctuations of the difference between price and return are given
by \be\label{paps} d(t,t')=\lim_{N\to\infty}
N\ovl{\davg{[R(t)-p(t)][R(t')-p(t')]}} \ee (note that the above
quantity with the explicit pre-factor $N$ turns out to be of order
one). This is equivalent to \BE \fl
d(t,t')=\lim_{N\to\infty}\frac{1}{\Omega}\sum_\omega\ovl{\davg{\l[m(t)+\Rtilde^\omega-a^\omega(t)\r]\l[m(t')+\Rtilde^\omega-a^\omega(t')\r]}},
\EE where we made use of (\ref{bel}). Proceeding as before and
integrating by parts over $b^\omega(t)$ one sees that \be
d(t,t')=\lim_{N\to\infty}\frac{1}{\Omega\Gamma^2}\sum_\omega
\ovl{\davg{\widehat{b}^\omega(t) \widehat{b}^\omega(t')}}.\ee  Anticipating that in the limit $N\to\infty$ the behavior of
$d(t,t')$ will be dominated by the same saddle-point describing
$\ovl{\davg{Z[\bsy{\psi}]}}$ and at which 
$\bsy{m}=\bsy{\varphi}=0$ we have \BE\fl d(t,t')=\frac{1}{\Gamma^2}\int D\bsy{Q}
D\bsy{\widehat{Q}} D\bsy{K} D\bsy{\widehat{K}} D\bsy{L}
D\bsy{\widehat{L}} D\bsy{\widehat{m}}
D\bsy{\widehat{\varphi}} ~
e^{N\l(\Psi+\Upsilon\r)+(\Omega-1)\Phi/\alpha+\Or(\sqrt{N})}\nonumber \\
\times\int Da D\widehat{a} Db
D\widehat{b}~\widehat{b}(t)\widehat{b}(t')\exp\l[\ii\sum_t\l[a(t)\widehat{a}(t)+b(t)\widehat{b}(t)+
\Gamma a(t)b(t)\r]\r]\\
\times\exp\l[-\frac{1}{2}\sum_{t,t'}\l[\Gamma^2\lambda^2
b(t)b(t')+Q(t,t')\widehat{a}(t)\widehat{a}(t')+
2K(t,t')\widehat{a}(t)\widehat{b}(t')\r]\r].\nonumber \EE Integrating over $a$, $\widehat{a}$ and $b$ we find \be
d(t,t')=\Lambda(t,t')/\Gamma^2=\l[(\bsy{1}+\Gamma
\bsy{G})^{-1}(\lambda^2\bsy{E}+\bsy{Q})(\bsy{1}+\Gamma
\bsy{G}^T)^{-1}\r](t,t'), \ee so that indeed $d(t,t')$ as defined above
is of order one. $\bsy{E}$ is the matrix with all
entries equal to one.

Following similar steps one can calculate the average price-return
difference, namely \be\label{papapa} A(t)=\lim_{N\to\infty} \sqrt{N}~
\ovl{\davg{[R(t)-p(t)]}}. \ee One finds \BE\fl
A(t)=\frac{1}{\Gamma}\int D\bsy{Q} D\bsy{\widehat{Q}} D\bsy{K}
D\bsy{\widehat{K}} D\bsy{L} D\bsy{\widehat{L}} 
D\bsy{\widehat{m}} D\bsy{\widehat{\varphi}} ~
e^{N\l(\Psi+\Upsilon\r)+(\Omega-1)\Phi/\alpha+\Or(\sqrt{N})}\nonumber \\
\int Da D\widehat{a} Db
D\widehat{b}~\widehat{b}(t)\exp\l[\ii\sum_t\l[a(t)\widehat{a}(t)+b(t)\widehat{b}(t)+
\Gamma a(t)b(t)\r]\r]\\
\times\exp\l[-\frac{1}{2}\sum_{t,t'}\l[\Gamma^2\lambda^2
b(t)b(t')+Q(t,t')\widehat{a}(t)\widehat{a}(t')+
2K(t,t')\widehat{a}(t)\widehat{b}(t')\r]\r].\nonumber \EE This
integral vanishes due to symmetry, hence $A(t)$ is a zero-average process
with temporal correlations $d(t,\tp)$ of order one.


\section{Ergodic steady states}
\subsection{General considerations}
Due to the presence of the coloured noise $\xi(t)$ and of the retarded
self-interaction in the effective process (\ref{eq:effprocess}) it is
in general not feasible to solve the self-consistent system
$\{$(\ref{pops}),(\ref{qu}),(\ref{kapa})$\}$ analytically for all
times $t, t'$. We shall therefore restrict ourselves to studying the
ergodic steady states of the effective-agent problem. These are time-translation invariant solutions, \be
\lim_{t\to\infty} Q(t+\tau,t)=Q(\tau),~~~
\lim_{t\to\infty} G(t+\tau,t)=G(\tau),~~~
\lim_{t\to\infty} \widehat{\varphi}(t)=\varphi,\nonumber \ee without
long-term memory, \be \lim_{t\to\infty} G(t,t')=0~~~~~~~\, \forall t'
~\mbox{finite}, \ee and with finite integrated response, \be
\lim_{t\to\infty}\sum_{\tau\leq t} G(\tau)=:\chi<\infty. \ee With these
Ans\"atze one performs a time-average of the effective
process (\ref{eq:effprocess}), leading to 
\be\label{eq:aveffprocess}
\widetilde{u}_\sigma=-\frac{1}{2}\varphi-\frac{1}{2}\Gamma\eta\alpha
z_\sigma-\frac{\alpha\Gamma\sigma/2}{1+\Gamma\chi}y+
\frac{\sqrt{\alpha}}{2}\sigma\xi. \ee 
Here, we have introduced $\widetilde{u}_\sigma=\lim_{t\to\infty}u_\sigma(t)/t$ (roughly representing the `velocity' with which the propensities grow in time), as well as the static variables
\be\fl
z_\sigma=\lim_{t\to\infty}\frac{1}{t}\sum_{t'\leq t}z_\sigma(t'),~~~~~~~
y=\lim_{t\to\infty}\frac{1}{t}\sum_{t'\leq t}y(t'),~~~~~~~
\xi=\lim_{t\to\infty}\frac{1}{t}\sum_{t'\leq t}\xi(t'). \ee $h_\sigma(t)$ has been set to zero. Note that $\xi$
is (static) Gaussian noise of zero mean and variance \be\label{eq:xivariance}
\avg{\xi^2}=\frac{\Gamma^2\l(\lambda^2+q\r)}{(1+\Gamma\chi)^2},  \ee
and that the $z_\sigma, \sigma\in\{-1,1\}$ as well as $y$ are stochastic variables, coupled to the value of $\xi$ via Eq. (\ref{eq:aveffprocess}).
The parameter \be q=\lim_{t\to\infty}\frac{1}{t}\sum_{\tau\leq t}
Q(\tau) \ee represents the persistent part of the correlation function
which, together with $\chi$ and $\varphi$, is to be determined self-consistently from \be \avg{x}=\Rbar,~~~~~~~ q=\avg{y^2},~~~~~~~
\chi=\frac{1}{\sqrt{\alpha}}\avg{\frac{\partial y}{\partial\xi}}. 
\label{saddol} \ee
Note that, up to a pre-factor, the noise acts effectively as an
external field, so that $\chi$ can be expressed in terms of a
derivative with respect to $\xi$.  Recalling (\ref{defs}) one realises
that $q$ is indeed the observable defined in (\ref{eq:qdef}).

In the stationary state the matrix $d(t,\tp)$ will also be time-translation invariant, and similar to the MG one finds 
\be\label{aitsch}
H=\alpha\lim_{\tau\to\infty} d(\tau)=\alpha\frac{\lambda^2+q}{(1+\Gamma\chi)^2}.
\ee
The magnitude of the fluctuations of the price around its temporal mean is given by 
\be
\delta p^2\equiv \alpha^{-1}\sum_\omega \avg{(p-\avg{p|\omega})^2|\omega}=d(\tau=0)-H/\alpha.
\ee
For this quantity one has the exact relation
\be
\delta p^2=\l[(\bsy{1}+\Gamma
\bsy{G})^{-1}(\lambda^2\bsy{E}+\bsy{Q})(\bsy{1}+\Gamma
\bsy{G}^T)^{-1}\r](0)-H/\alpha.
\ee
A further evaluation of $\delta p^2$ hence requires in principle the full knowledge of the transient contributions to the correlation and response functions. While the analogous quantity in the MG can be well approximated in terms of persistent order parameters, such an estimate appears much more subtle here due to an explicit dependence of the fluctuations on the learning rate. Defining $\widetilde Q(t,\tp)=Q(t,\tp)-q$ one has
\be
\delta p^2=\l[(\bsy{1}+\Gamma
\bsy{G})^{-1}\bsy{\widetilde Q}(\bsy{1}+\Gamma
\bsy{G}^T)^{-1}\r](0),
\ee
where $\bsy{\widetilde Q}$ is a measure of the fluctuations of the variables $y_i(t)$ around their temporal averages, and depends on the learning rate. For $\Gamma\to 0$ no fluctuations are present, so that $\delta p^2\to 0$, as pointed out in \cite{Bergetal}. In general, the right-hand-side is an increasing function of $\Gamma$, which is difficult to express in terms of persistent order parameters within the present setup. We choose here to focus on the asymptotics of the model.

In order to proceed with the analysis of (\ref{eq:aveffprocess}) one inspects the behaviour of the $\{u_{i\sigma}(t)\}$ in numerical simulations, and formulates suitable Ans\"atze for $\widetilde{u}_\sigma$, corresponding to different types of solutions observed in simulations. For purposes of clarity, we will treat the cases $\eta=0$ and $\eta>0$ separately in the following.

\subsection{$\eta=0$}

Simulations of the model without impact-correction reveal the
existence of three distinct phases, two ergodic ones and a third with
anomalous response:
\begin{itemize}
\item For \underline{large $\alpha$} greater than a critical value
  $\alpha^\star$ both propensities $u_{i\sigma}$ are positive on average
  and remain finite in the steady state for all agents. Each agent
  asymptotically invests finite amounts under both signals. The
  corresponding effective agent has $\widetilde{u}_\sigma=0$ for both
  $\sigma\in\{-1,1\}$. We shall refer to this phase as the
  `$2$-phase', indicating that all agents invest under both signals.
\item For \underline{intermediate $\alpha$} agents are divided into two
  groups. Some agents do not invest under either signal, both of their
  propensities decrease linearly with time, so that
  $\widetilde{u}_\sigma<0$ for both $\sigma\in\{-1,1\}$. Each of the
  remaining agents invest under one signal (say $\sigma_i$ for player
  $i$) but not under the opposite signal $-\sigma_i$. For such agents,
  the propensity $u_{i\sigma_i}(t)$ is positive and remains finite
  asymptotically, while the other one decreases linearly in time. This corresponds to trajectories of the effective process with $\widetilde{u}_{\sigma}=0$ for one value $\sigma\in\{-1,1\}$ and
  $\widetilde{u}_{-\sigma}<0$. We will refer to this phase as
  the `$1+0$-phase', indicating that some players do not invest under
  either signal, while others invest under precisely one signal.
\item  For \underline{low $\alpha$} less than a critical value $\alpha_c$ one finds that the macroscopic order parameters of the steady state depend on initial conditions, see Fig. \ref{fig:batcheqonline}. Hence the dynamics is
non-ergodic, and we expect the integrated response $\chi$ to be infinite in this regime. We will label this phase by {\bf NE} (non-ergodic) in the following.
\end{itemize}

\begin{figure}[t]
\vspace*{-6mm} \hspace*{35mm} \setlength{\unitlength}{1.3mm}
\begin{picture}(120,55)
\put(-8,-0){\epsfysize=50\unitlength\epsfbox{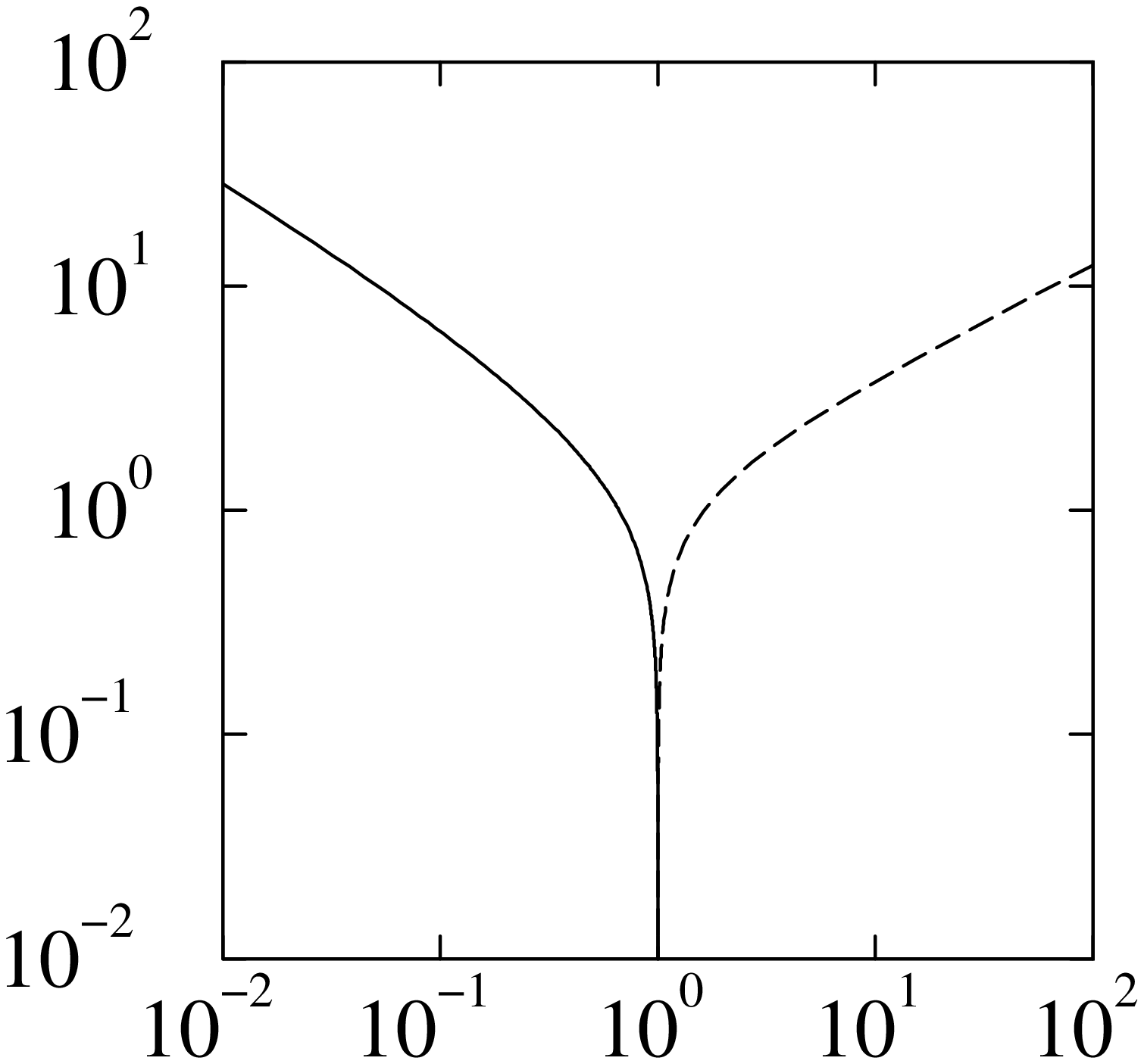}}
\put(-20,29){\Large $\lambda/\overline{R}$} \put(22,-5){\Large $\alpha$}
\put(30,16){\large\bf 2} \put(20,40){\large\bf 1+0}
\put(10,16){\large\bf NE}
\end{picture}
\vspace*{-0mm} \caption{Phase diagram of the model without impact-correction. Solid line indicates the ergodic/non-ergodic ({\bf NE}) transition at $\alpha_c$, given by Eq. (\ref{eq:alphac}), dashed line is the transition between the $2$- and the $1+0$-phases at $\alpha^\star$, see Eq. (\ref{eq:alphastar}).} \label{fig:phasediagram}
\end{figure}
 It will turn out that the relevant control parameters for the phase behaviour of the model are given by $\alpha$ and the ratio $\lambda/\Rbar$ of the mean return over its standard deviation, the learning rate $\Gamma$ has no influence on the persistent order parameters in the stationary states, but only on the transients of the dynamics. The resulting phase diagram in the $(\alpha,\lambda/\Rbar)$-plane is depicted in Fig. \ref{fig:phasediagram}

We will now study the two ergodic  phases separately, starting with the phase at intermediate values of $\alpha$, and will compute the persistent order parameters in the two ergodic phases, as well as the boundaries $\alpha^\star$ and $\alpha_c$ separating the three regimes.

\subsubsection{The $1+0$ phase at intermediate $\alpha$}

Agents who do not invest under either signal correspond to solutions
of the effective agent process with \be
\widetilde{u}_\sigma=-u_\sigma<0,~~~~~~~z_\sigma=y=0,~~~~~~~\sigma\in\{-1,1\}.
\ee While the specific values of the $u_\sigma, \sigma\in\{-1,1\}$ will
play no role for the further analysis, we would like to stress that
different realisations of the effective process (that is different
realisations of the noise $\xi$) can in general lead to different
values for the $u_\sigma, \sigma\in\{-1,1\}$ and for the $z_\sigma$.  For agents who
invest under one signal, but not under the other, we will inspect
solutions of the effective process of the following two types: \BE
(\widetilde{u}_+, \widetilde{u}_-)=(0,-u)~~~~~~~
(z_+,z_-)=(2y,0)~~~~~~~~~y>0, \label{posuno}\\
(\widetilde{u}_+, \widetilde{u}_-)=(-u,0)~~~~~~~
(z_+,z_-)=(0,-2y)~~~~~~~y<0. \label{posdue} \EE The former corresponds
to agents who invest under signal $\sigma=1$, the latter to agents who
play upon receiving $\sigma=-1$. In both cases, $u$ takes a positive
value, which again in principle may vary for different realisations of
the effective process.  In either case summing the two relations
(\ref{eq:aveffprocess}) (with $\eta=0$) leads to \be u=\varphi
\label{io}. \ee
In particular $u$ takes the same value for all (effective) agents who invest under exactly one signal.  Self-consistency demands $\varphi>0$ as we have assumed above that $u$ is positive. Taking the difference of the two equations of (\ref{eq:aveffprocess}) we find  \be \label{ioo}
\frac{\alpha\Gamma|y|}{1+\Gamma\chi}=-\varphi\pm\sqrt{\alpha}\xi, \ee
where the plus signs describes the case $y>0$ while the minus sign
holds when $y<0$. Setting $\xi^\star=\varphi/\sqrt{\alpha}$ one sees
that the former case is realized when $\xi>\xi^\star$, the latter when
$\xi<-\xi^\star$. For $|\xi|<\xi^\star$ no solution with
$|y|>0$ is possible and the corresponding effective agent never
invests, as discussed above. We conclude that the physical interpretation
of $\varphi$ is closely related to the relative weight of the two
types of agents. Indeed, the probability that an agent is inactive,
that is the fraction of agents who do not invest under either signal, is given by \be
\phi_0=\avg{\theta\l(\xi^\star-|\xi|\r)}=\erf\l(\frac{\varphi}{
\sqrt{2\alpha\gamma}}\r), \ee where $\gamma=\avg{\xi^2}$, see
(\ref{eq:xivariance}). The persistent order parameters $\varphi$, $q$
and $\chi$ are obtained from the self-consistency relations \be
\avg{\frac{z_++z_-}{2}}=\Rbar,~~~~~~~ q=\avg{y^2},~~~~~~~
\chi=\frac{1}{\sqrt{\alpha}} \avg{ \frac{\partial y}{\partial\xi}}, \ee
which may be written as
 \BE \frac{\Gamma\Rbar\sqrt{\alpha}}{1+\Gamma\chi}=
\avg{(\xi-\xi^\star)\theta(\xi-\xi^\star)}+
\avg{(-\xi-\xi^\star)\theta(-\xi-\xi^\star)},\nonumber\\
\frac{\alpha\Gamma q}{\l(1+\Gamma\chi\r)^2}=
\avg{(\xi-\xi^\star)^2\theta(\xi-\xi^\star)}+
\avg{(-\xi-\xi^\star)^2\theta(-\xi-\xi^\star)},\\
\frac{\alpha\Gamma\chi}{1+\Gamma\chi}= \avg{\theta(\xi-\xi^\star)}+
\avg{\theta(-\xi-\xi^\star)}.\nonumber \EE After carrying out the
remaining integrations over $\xi$ one finds \BE \frac{\Gamma\Rbar}{1+\Gamma\chi}
&=&2\sqrt{\frac{\gamma}{2\pi\alpha}}\exp \left(-\frac{\varphi^2}{2\alpha\gamma}\right)-\frac{\varphi}{\alpha}~\erfc\left(\frac{\varphi}{\sqrt{2\alpha\gamma}}\right),\nonumber\\
 \frac{\alpha\Gamma^2 q}{\l(1+\Gamma\chi\r)^2}
&=&\l(\gamma+\frac{\varphi^2}{\alpha}\r)\erfc\left(\frac{\varphi}{\sqrt{2\alpha\gamma}}\right)-2\varphi\sqrt{\frac{\gamma}{2\pi\alpha}}\exp\left(-\frac{\varphi^2}{2\alpha\gamma}\right),\label{eq:im2}\\
 \frac{\alpha\Gamma\chi}{1+\Gamma\chi}
&=&1-\erf\left(\frac{\varphi}{\sqrt{2\alpha\gamma}}\right).\nonumber \EE
This coupled system of non-linear equations is easily solved numerically (for example using Newton-Raphson methods), and the order parameters $q, \chi$ and $\varphi$ may be obtained as functions of $\alpha$ for any fixed values of the model parameters $\Rbar$, $\lambda$ and $\Gamma$. The dependence of these persistent order parameters on the learning rate $\Gamma$ can be understood by an inspection of (\ref{eq:im2}). One finds that the solution is $\Gamma$-independent when expressed in terms of the re-scaled variables $\{q,\Gamma\chi,\varphi/\Gamma\}$. $H$ can in turn be obtained from (\ref{aitsch}).

This solution is valid self-consistently as long as $\chi$ turns out
to be finite, and as long as $\varphi$ comes out positive. The point
at which the latter condition breaks down is easily determined upon
setting $\varphi=0$ in the above coupled set of equations. After some
algebra one finds that this occurs at \be\label{eq:alphastar}
\alpha=\alpha^\star=1+\frac{2}{\pi}\frac{\lambda^2}{\Rbar^2}, \ee which
coincides with the static result \cite{Bergetal}.  The onset of
anomalous response, i.e. the point $\alpha_c$ at which $\chi$ diverges
at fixed values of $\lambda$, $\Gamma$ and $\Rbar$ is found to be
determined by the condition\footnote{Note that a very similar
  condition $\alpha_c=1-\phi(\alpha_c)$ has been found in the context
  of the Minority Game \cite{HeimCool01}. There, $\phi$ denotes the
  fraction of so-called frozen agents.}\be\label{eq:alphac}
\alpha_c=1-\phi_0(\alpha_c).  \ee $\alpha_c$ is obtained as
$\alpha_c=\mbox{erfc}(\zeta_c)$ where $\zeta_c$ is the root of
\be
\frac{\Rbar}{\lambda}=\frac{e^{-\zeta^2}/\sqrt{\pi}-\zeta\mbox{erfc}(\zeta)}{\sqrt{\zeta
e^{-\zeta^2}/\sqrt{\pi}-\zeta^2\mbox{erfc}(\zeta)}}.
\ee
Note that due to Eq. (\ref{aitsch}) $H$ vanishes at the
point of diverging $\chi$. Hence the dynamical phase transition
between the ergodic and non-ergodic regimes coincides with the
transition between efficient ($H=0$) and non-efficient ($H>0$) phases
observed in \cite{Bergetal}.  One finds numerically that
$\alpha_c<\alpha^\star$ for all fixed values of the parameters $\lambda,
\Gamma, \Rbar$, so that we conclude that the $1+0$-phase is physically
realised for intermediate values of $\alpha\in[\alpha_c,\alpha^\star]$.

\subsubsection{The $2$-phase at $\alpha>\alpha^\star$.}

Here we set $\widetilde{u}_\sigma=0$ for both $\sigma\in\{-1,1\}$ in (\ref{eq:aveffprocess}). Summing the resulting expressions for $\sigma=1$ and $\sigma=-1$ one immediately finds $\varphi=0$ for $\eta=0$. Taking the difference, instead, yields \be
\frac{y\Gamma\sqrt{\alpha}}{1+\Gamma\chi}=\xi. \ee This in turn implies that \be
q=\avg{y^2}=\frac{\avg{\xi^2}\l(1+\Gamma\chi\r)^2}{\alpha\Gamma^2}=
\frac{\lambda^2+q}{\alpha}, \ee from which we can directly read off the value of $q$ in the $2$-phase:   
\be q=\frac{\lambda^2}{\alpha-1},
\label{eq:q2} 
\ee in agreement with the corresponding static result given in \cite{Bergetal}. For the
susceptibility one obtains \be\label{eq:chi2} \chi=\frac{1}{\sqrt{\alpha}}\avg{\frac{\partial y}{\partial\xi}} =\frac{1}{\Gamma(\alpha-1)}.
\label{narasta}\ee Eqs. (\ref{eq:q2}) and (\ref{eq:chi2}) along with our result $\varphi=0$ completely describe the persistent order parameters in the ergodic
steady states at $\alpha>\alpha^\star$. Note that $\alpha^\star>1$ by virtue of (\ref{eq:alphastar}), so that no singularities occur in the $2$-phase.
Using Eq. (\ref{aitsch}) $H$ is given by
\be
H=\lambda^2(\alpha-1) 
\ee 
for $\alpha>\alpha^\star$.

\subsubsection{Comparison with simulations}
\begin{figure}[t]
\vspace*{1mm}
\begin{tabular}{cc}
\epsfxsize=75mm  \epsffile{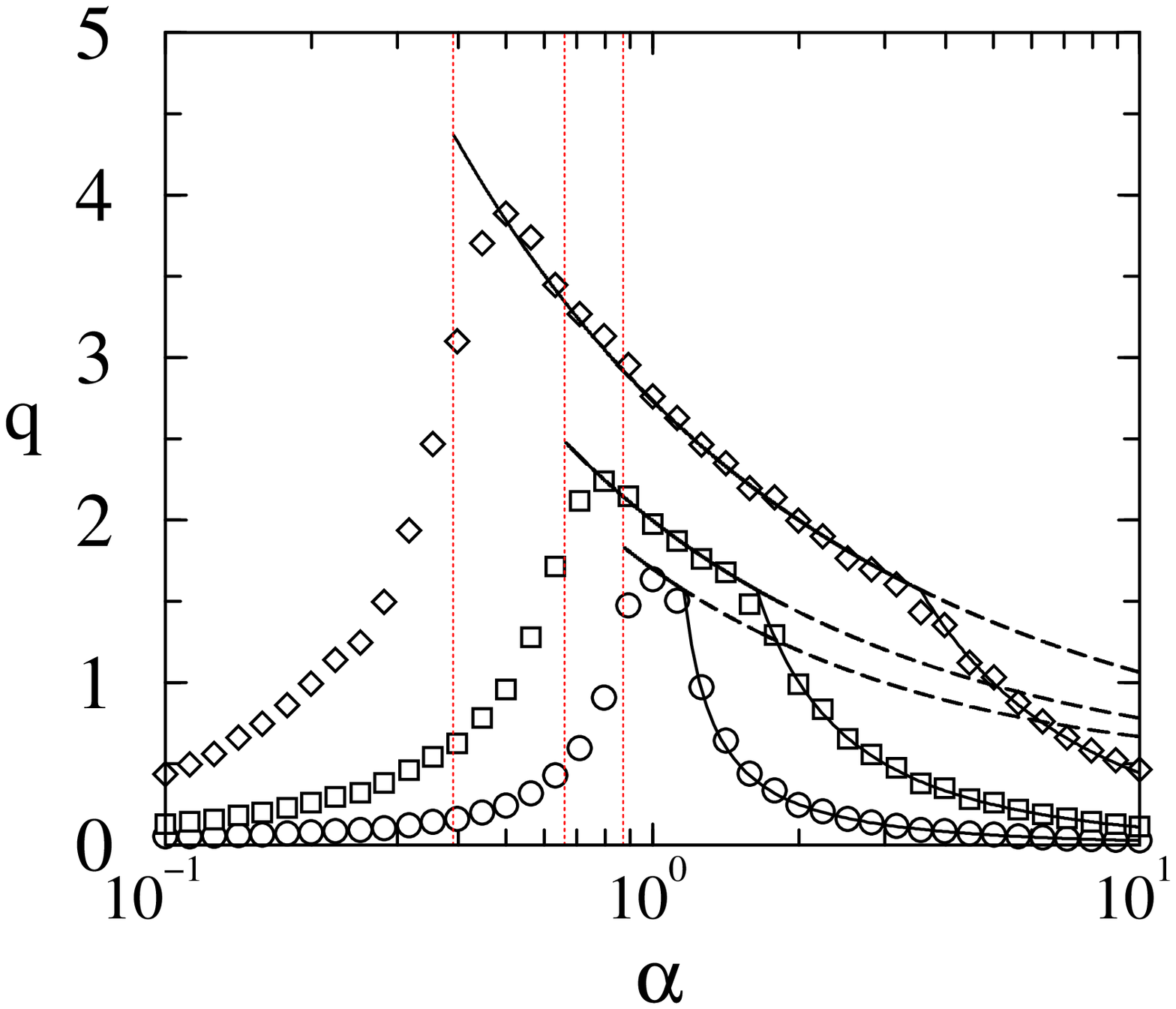} ~~&~~
\epsfxsize=72mm  \epsffile{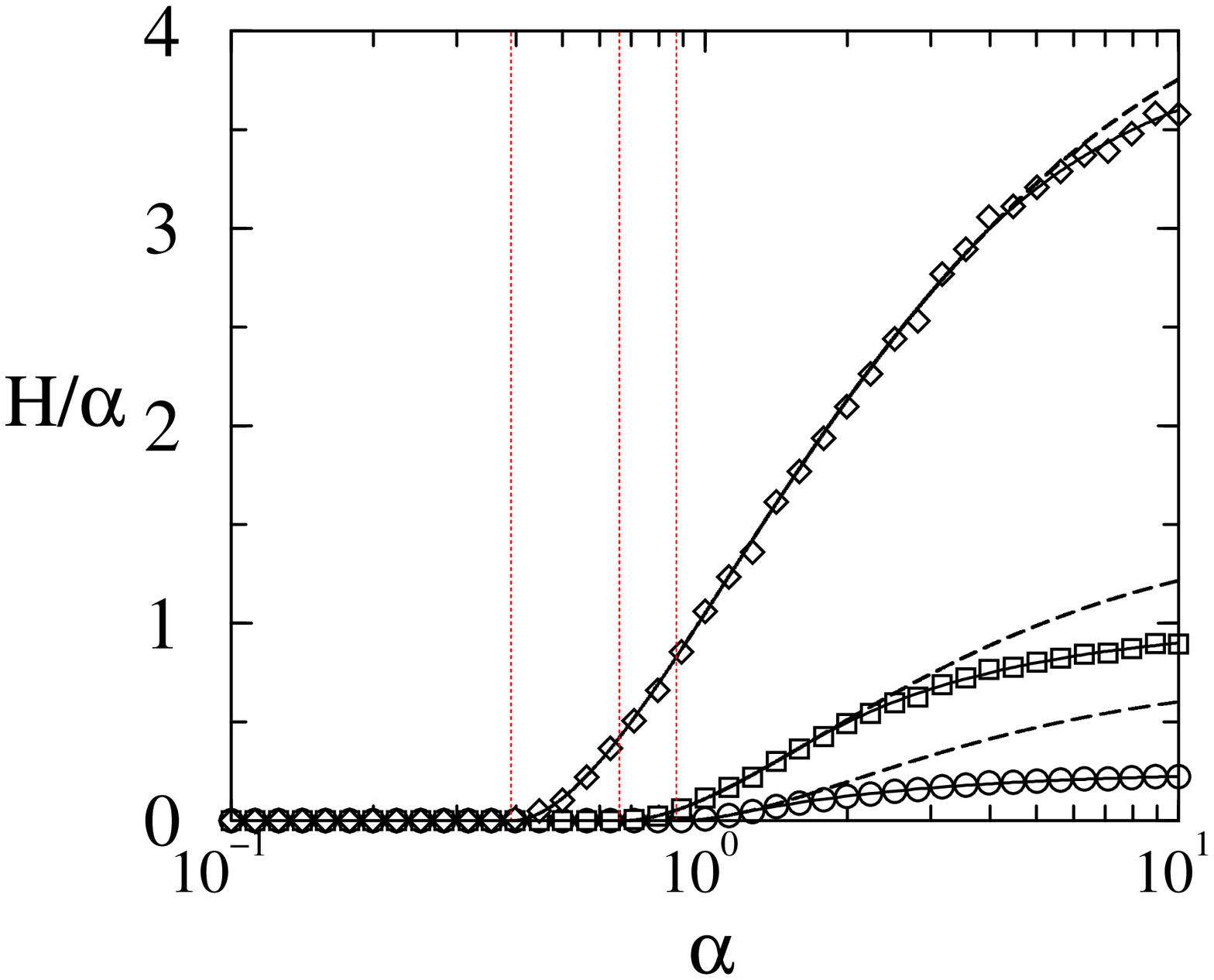}
\end{tabular}
\vspace*{1mm} \caption{Order parameters $q$ and $H/\alpha$ as functions of $\alpha$ for fixed $\eta=0$. Markers are from simulations of the on-line model for $\lambda=0.5$ (circles), $\lambda=1$ (squares) and $\lambda=2$ (diamonds), started from unbiased initial conditions. We set $\Gamma=1$ for all three curves. The solid lines are the predictions of the analytical theory and have been continued as dashed lines into phases where they are no longer valid. The vertical lines indicate the analytically obtained locations of the ergodic/non-ergodic phase transition at $\alpha_c$ for the three different values of $\lambda=0.5,1,2$ (from right to left).} \label{fig:qh_vs_alpha_eta_0}
\end{figure}

\begin{figure}
\vspace*{1mm}
\begin{center}
\epsfxsize=85mm  \epsffile{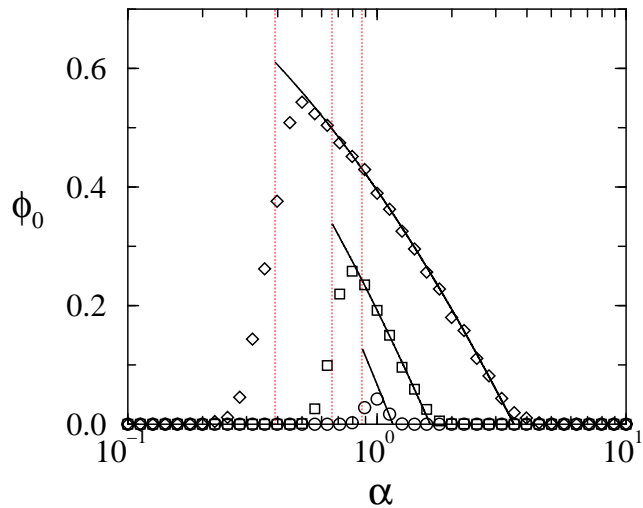}\\
\end{center}
\vspace*{1mm} \caption{Fraction $\phi_0$ of agents who do not invest under either signal for fixed $\eta=0$.  Markers are from simulations of the on-line model for $\lambda=0.5$ (circles), $\lambda=1$ (squares) and $\lambda=2$ (diamonds), started from unbiased initial conditions. We set $\Gamma=1$ for all three curves. The solid lines are the predictions of the analytical theory in the phase of intermediate $\alpha$. Outside this phase $\phi_0$ is predicted to be zero. The vertical lines indicate the analytically obtained locations of the ergodic/non-ergodic phase transition at $\alpha_c$ for the three different values of $\lambda=0.5,1,2$ (from right to left).
\label{fig:phi0_eta_0}}
\end{figure}

We have tested our theoretical predictions for the game without impact-correction against numerical simulations. Results for $H$ and $q$ are presented in Fig. \ref{fig:qh_vs_alpha_eta_0}, while Fig. \ref{fig:phi0_eta_0} shows the behaviour of $\phi_0$. All simulations are performed on the on-line update rules (given by Eq. (\ref{eq:onlineupdate})) with $\alpha N^2=10^4$. Measurements are taken over $50000$ time-steps preceded by $150000$ equilibration steps. All data presented are averages over $100$ samples of the disorder.  The learning rate is kept fixed at $\Gamma=1$ (we have verified the independence of $q$ and $H$ of the learning rate in separate simulations). The figures demonstrate very good agreement of the theoretical predictions with the numerical data, modulo finite-size effects close to the transition points. We observe that $q$ is a decreasing function of $\alpha$ in the two ergodic phases, with a cusp at the transition point between the $1+0$- and the $2$-phase at $\alpha=\alpha^\star$. The breakdown of ergodicity below $\alpha_c$ can be illustrated by starting the dynamics from differently biased initial propensities. While the macroscopic order parameter $q$ is insensitive to initial conditions above $\alpha_c$, the starting point becomes relevant in the non-ergodic phase, as shown in Fig. \ref{fig:batcheqonline}. 

\subsection{$\eta>0$}

For $\eta>0$ the situation is slightly more complicated. One observes
at all $\alpha$ that agents are divided in two classes: those who
trade under both signals and those who trade at most under one
signal. Let us discuss this scenario in detail. The former have \be
\widetilde{u}_\sigma=0,~~~~~~~z_\sigma>0,~~~~~~~\sigma\in\{-1,1\}. \ee
Notice that for these agents $x=(z_++z_-)/2>|y|=|(z_+-z_-)/2|$. For the latter
one has instead (\ref{posuno}) and (\ref{posdue}) as before. For them,
$x=y$ when $y>0$ and $x=-y$ when $y<0$.

\begin{figure}[t]
\vspace*{1mm}
\begin{tabular}{cc}
\epsfxsize=75mm  \epsffile{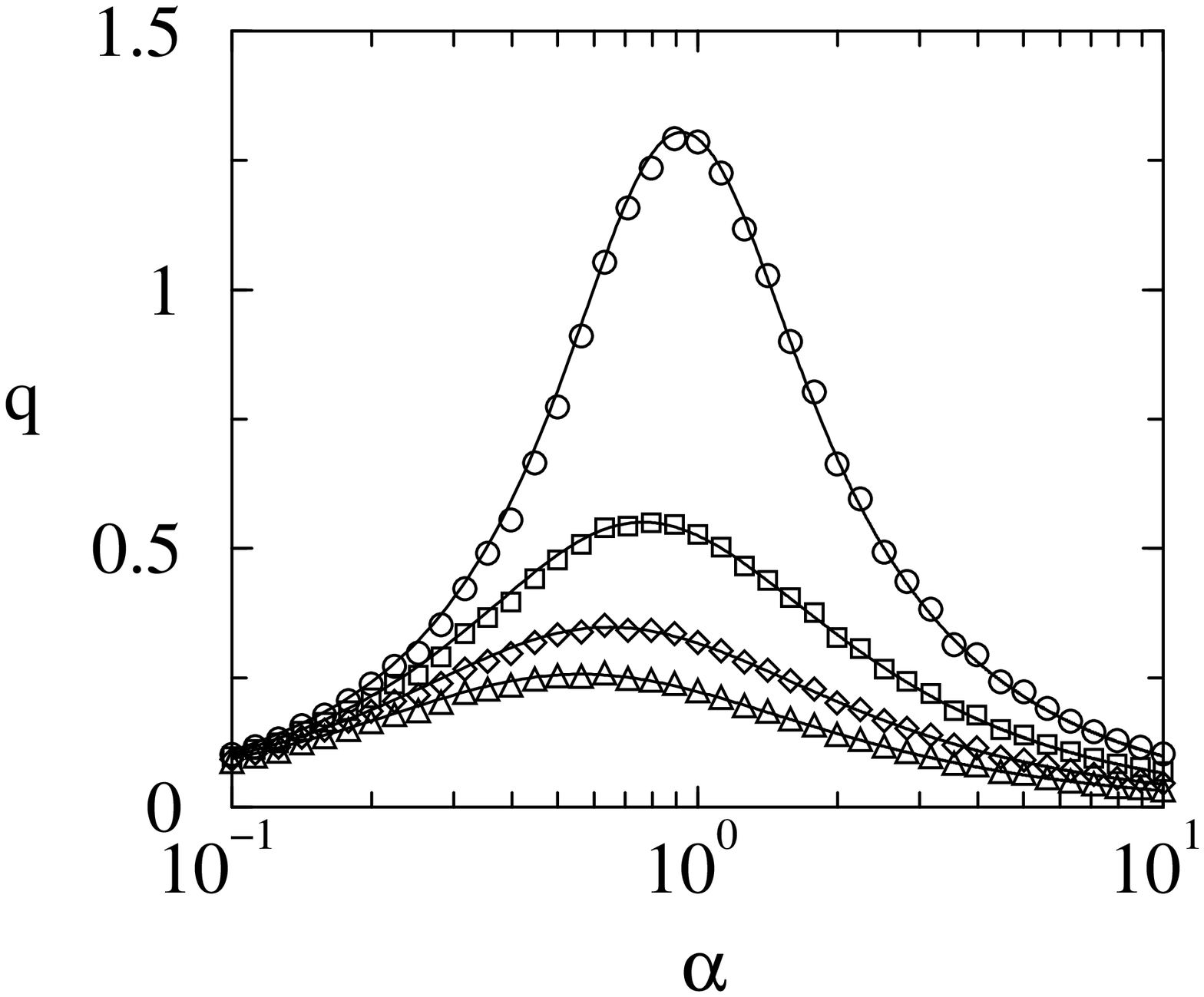} ~~&~~
\epsfxsize=72mm  \epsffile{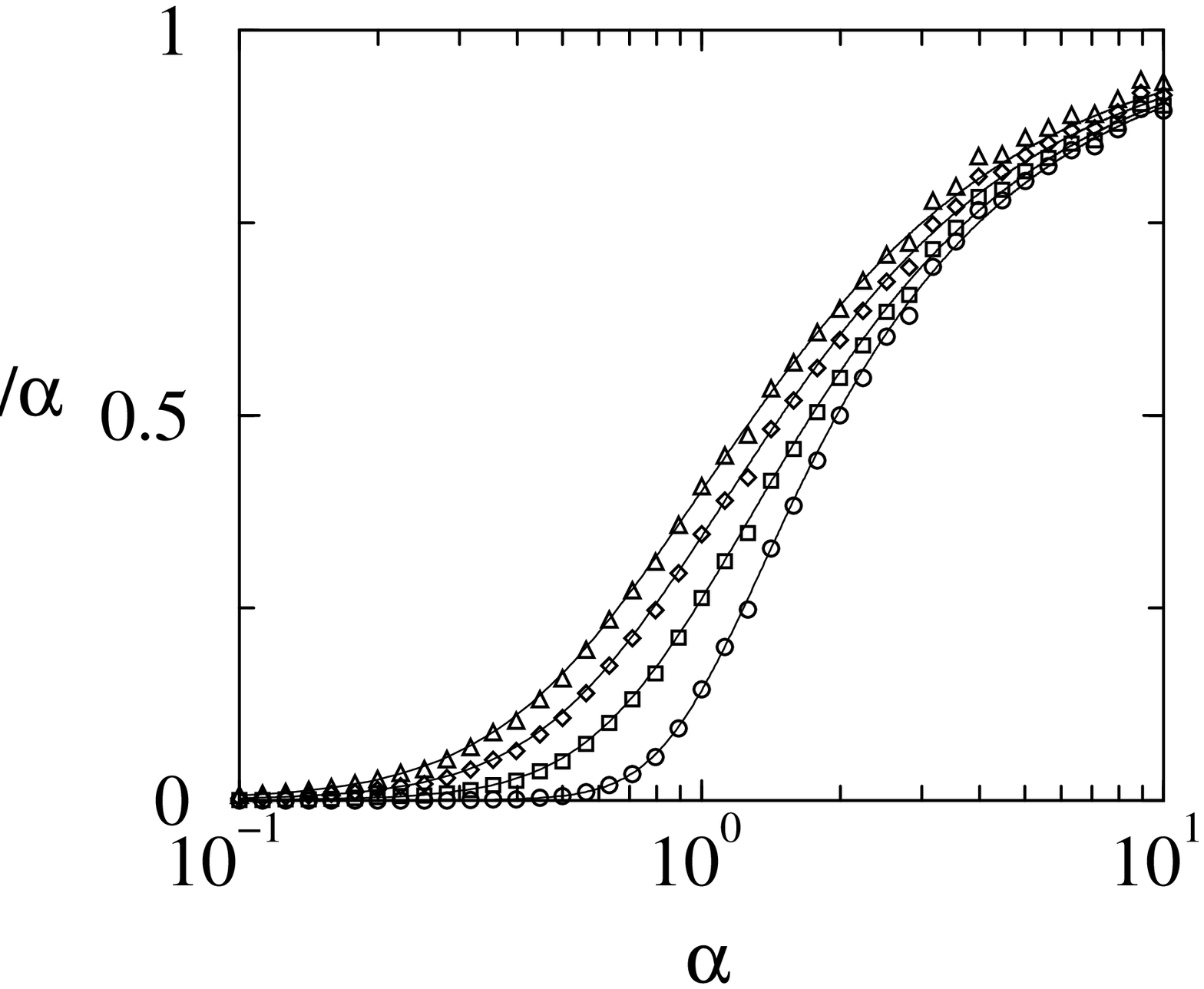}
\end{tabular}
\vspace*{4mm} \caption{Order parameters $q$ and $H/\alpha$ for $\eta>0$ as functions of $\alpha$ at fixed values of $\Gamma=\lambda=1$. Markers are from simulations of the on-line model for $\eta=0.05$ (circles), $\eta=0.25$ (squares), $\eta=0.5$ (diamonds) and $\eta=0.75$ (triangles), started from unbiased initial conditions. The solid lines are the predictions of the analytical theory in the $1+2$ phase.} \label{fig:qh_vs_alpha_different_eta}
\end{figure}

\begin{figure}
\vspace*{1mm}
\begin{center}
\epsfxsize=85mm  \epsffile{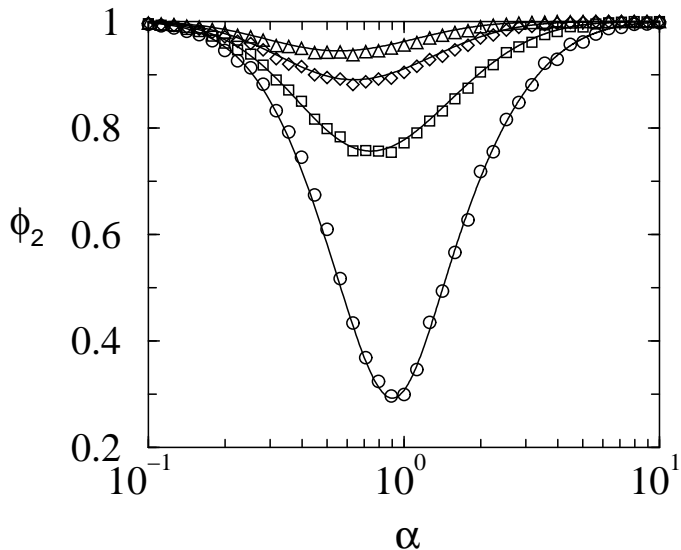}\\
\end{center}
\vspace*{4mm} \caption{Fraction $\phi_2$ of agents who invest under both signals, shown  for different values of $\eta>0$.  Markers are from simulations of the on-line model for $\eta=0.05$ (circles), $\eta=0.25$ (squares), $\eta=0.5$ (diamonds) and $\eta=0.75$ (triangles), started from unbiased initial conditions and with $\Gamma=1$, $\lambda=1$. The solid lines are the predictions of the analytical theory in the $1+2$ phase. \label{fig:phi2_different_eta}}
\end{figure}

Let us start with the traders who always invest. Summing equations
(\ref{eq:aveffprocess}) with non-zero $\eta$ and
$\widetilde{u}_\sigma=0$ one gets \be 
\label{varp} \varphi=-\Gamma\eta\alpha x, \ee
whereas taking the difference gives \be
y=\frac{\xi}{\sqrt{\alpha}\Gamma\l(\eta+\frac{1}{1+\Gamma\chi}\r)}. \ee
The requirement $|y|<x$ now translates into the condition\be
|\xi|<-\frac{\varphi}{\eta\sqrt{\alpha}}\l(\eta+\frac{1}{1+\Gamma\chi}
\r) \ee on the effective noise. Note that $\varphi<0$ by virtue of
(\ref{varp}). Turning to agents who trade under one signal only,
summing equations (\ref{eq:aveffprocess}) with non-zero $\eta$ one
finds that (\ref{io}) and (\ref{ioo}) generalize to  \BE
u=\varphi+\Gamma\eta\alpha|y|, \label{iooo}\\
\alpha\Gamma|y|\l(\eta+\frac{1}{1+\Gamma\chi}\r)\mp\sqrt{\alpha}\xi=-u,
\label{ioooo} \EE where the minus (resp. plus) sign holds for agents with $y>0$
(resp. $y<0$). Let us focus on agents with $y>0$. Combining
(\ref{iooo}) and (\ref{ioooo}) one gets \be \label{mommo}
y=\frac{-\varphi+\sqrt{\alpha}
  \xi}{\alpha\Gamma\l(2\eta+\frac{1}{1+\Gamma\chi}\r)}. \ee On the
other hand, since $u>0$ we must have \be \xi>\sqrt{\alpha}\Gamma
y\l(\eta+\frac{1}{1+\Gamma\chi}\r), \ee which via (\ref{mommo})
becomes \be \xi>-\frac{\varphi}{\eta\sqrt{\alpha}}
\l(\eta+\frac{1}{1+\Gamma\chi}\r). \ee Similarly one finds that the
solution with $y<0$ corresponds to \be
\xi<\frac{\varphi}{\eta\sqrt{\alpha}}\l(\eta+\frac{1}{1+\Gamma\chi}\r).
\ee Hence, defining
$\ovl{\xi}=-\frac{\varphi}{\eta\sqrt{\alpha}}\l(\eta+
\frac{1}{1+\Gamma\chi}\r)$, the fraction of players who invest under both signals can be written as 
\be
\phi_2=\avg{\theta(\ovl{\xi}-|\xi|)},
\ee 
where $\avg{\dots}$ is again an average over $\xi$ (with variance given by (\ref{eq:xivariance})). The saddle-point conditions (\ref{saddol})
take the form: \BE \fl
\sqrt{\alpha}\Gamma\Rbar=-\frac{\varphi}{\eta\sqrt{\alpha}}
\avg{\theta(\ovl{\xi}-|\xi|)}+\frac{
  \avg{(\xi-\xi^\star)\theta(\xi-\ovl{\xi})}
  -\avg{(\xi+\xi^\star)\theta(-\ovl{\xi}-\xi)}}{2
  \eta+1/(1+\Gamma\chi)},\nonumber\\ \fl
\alpha\Gamma^2q=\frac{\avg{\xi^2\theta(\ovl{\xi}-\xi)}}{[\eta+
  1/(1+\Gamma\chi)]^2}+\frac{\avg{(\xi-\xi^\star)^2\theta(\xi-\ovl{\xi})}+
  \avg{(\xi+\xi^\star)^2\theta(-\xi-\ovl{\xi})}}{[2
  \eta+1/(1+\Gamma\chi)]^2},\label{weissichnich}\\ \fl
\alpha\Gamma\chi=\frac{\avg{\theta(\ovl{\xi}-\xi)}}{\eta+
  1/(1+\Gamma\chi)}+\frac{\avg{\theta(|\xi|-\ovl{\xi})}}{2
  \eta+1/(1+\Gamma\chi)},\nonumber \EE where as before
  $\xi^\star=\varphi/\sqrt{\alpha}$. We do not report the lengthy
  expressions one obtains after the averages over $\xi$ are carried
  out, but would like to stress that one readily checks that these
  equations do not allow for a diverging susceptibility at any finite
  $\alpha$.  Therefore the model with impact-correction does not
  exhibit anomalous response, in contrast with the game at
  $\eta=0$. From (\ref{weissichnich}) the order parameters $\varphi$,
  $q$ and $\chi$ can be obtained numerically as functions of $\alpha$
  for any fixed values of $\eta>0$, $\lambda$ and $\Gamma$, and near perfect agreement is found with numerical simulations, see Fig. \ref{fig:qh_vs_alpha_different_eta} and \ref{fig:phi2_different_eta}. Observing no systematic deviations from the theory, we have no reason to suspect an onset of long-term memory at finite $\chi$, and hence the physical picture of the present model at $\eta>0$ appears different from the behaviour of the MG with impact correction \cite{HeimDeMa01}.

\section{Concluding remarks}
We have presented an analysis of the dynamics of a system of adaptive
agents with private asymmetric information, complementing and
extending the study of the statics of the model previously presented
in \cite{Bergetal}. To this end we have devised a batch version of the
original on-line update rules and observe no significant effects on
the persistent order parameters in the stationary states. This
demonstrates that the replacement of the on-line dynamics by
an information-averaged batch process as successfully performed in the
context of the MG can be extended to other models of interacting
agents. Path integral methods can then be used to turn the coupled
batch dynamics of the $N$ interacting agents into an effective
single-agent problem in the limit $N\to\infty$. From this effective
process the persistent order parameters in the different ergodic
stationary states as well as the phase diagram can be computed
exactly and in agreement with the static results obtained via replica
techniques. For the model without impact-correction ($\eta=0$) we
find three different phases, two ergodic ones and a phase with broken
ergodicity and dependence of the stationary macroscopic order
parameters on initial conditions. The location of the onset of
ergodicity breaking, $\alpha_c$, coincides with the location of the
transition between efficient and non-efficient phases identified in
\cite{Bergetal}. For sophisticated agents ($\eta>0$) only one phase is
present, and no ergodicity breaking occurs. The generating functional
approach also allows to address the issue of (average) fluctuations of
prices around the (quenched) returns, and an exact relation to the
dynamical order parameters can be drawn. The computation of $H$ does not require the knowledge of the transient contributions of the correlation and response functions, but only of their persistent parts (which we can compute exactly). On the other hand full solutions of the self-consistent effective problem for the functions $C(\tau)$ and $G(\tau)$ are needed to calculate the fluctuations of $p(t)$ around $R(t)$. While the corresponding fluctuations in the MG (the so-called volatility) can be well estimated in terms of persistent order parameters, similar approximations appear to be much more delicate in the present model. In addition we find that the volatilities of the batch and on-line models are different, which is not is not surprising as they depend on transients in $C$ and $G$. Moreover a dependence of the magnitude of fluctuations on the learning rate has been reported in \cite{Bergetal}. An analytical study of this dependence is beyond the scope of the present paper; it appears that these issues are more effectively tackled in suitably simplified versions of the model of \cite{Bergetal}, work on which is in progress.

In conclusion the dynamical mean field theory extensively used in the
context of the MG with common public information can be extended to
models of interacting agents with asymmetric non-uniform
information. The present model can up to now presumably at best be
seen as a most simplistic model of a financial market. Possible
extensions include models with heterogeneous learning rates
$\Gamma_i$, such models are of interest both from the mathematical
point of view as they would lead to an ensemble of effective processes
similar to \cite{DeMa03,GallSher05}, but would also allow to study the
interaction and relative success of agents with different abilities of
adaptation. In the same realm the individual wealth of the agents
could be taken into account, each varying in time according to the
performance of the agents. It might also be worthwhile studying the
influence of decision noise on the phase diagram. Another presumably
most interesting extension of the model would be one in which the
information available to the agents is not only asymmetric, but also
noisy. Finally, in the present model the `states of the world'
$\omega$ are drawn at random at each time step. With techniques now
available to study Minority Games with real histories \cite{Book2}, an
attempt might be made to replace this external random signal by
endogenously generated pieces of information relating to the previous
history of the market. The state $\omega(t)$ at a given time step $t$ might
for example encode the prices $p(t-1),p(t-2),\dots,p(t-M)$ of the
previous $M$ steps or relate to the history of differences
between prices and returns. Work along some of these lines is
currently in progress.

\section*{Acknowledgements}
This work was supported by the European Community's Human Potential
Programme under contract HPRN-CT-2002-00319, STIPCO, and by EVERGROW,
integrated project No. 1935 in the complex systems initiative of the
Future and Emerging Technologies directorate of the IST Priority, EU
Sixth Framework. ADM wishes to thank the Abdus Salam ICTP for
hospitality. The authors are grateful to J M Berg and M Marsili for
helpful discussions.



\section*{References}

\begin{thebibliography}{99}
\bibitem{Bergetal}Berg J, Marsili M, Rustichini A and Zecchina R 2001
  {\em J. Quant. Finance} {\bf 1(2)} 203
\bibitem{ChalZhan97}
Challet D and  Zhang Y-C 1997 \emph{Physica A} {\bf 246}  407
\bibitem{Book1}Challet D, Marsili M and Zhang Y-C 2005 {\em Minority
    Games} (Oxford University Press, Oxford UK)
\bibitem{Book2}Coolen ACC 2005 {\em The Mathematical Theory of
    Minority Games} (Oxford University Press, Oxford UK)
\bibitem{Book3}Johnson NF, Jefferies P and Hui PM 2003 {\it Financial
    market complexity} (Oxford University Press, Oxford UK)
\bibitem{Cava99}
Cavagna A 1999 {\em Phys. Rev. E} {\bf 59}  R3783

\bibitem{asymminfo} Akerlof G 1970 {\em Quarterly Journal of Economics} {\bf 84(3)} 488

\bibitem{ShapShub}
Shapley L and Shubik M 1977 {\em Journal of Political Economy} {\bf 85} 937
\bibitem{DeDom78}De Dominicis C 1978 {\em Phys. Rev. B} {\bf 18} 4913
\bibitem{MarsChalZecc00}
Marsili M, Challet D and Zecchina R 2000 {\em Physica A} {\bf 280}
522
\bibitem{HeimCool01}
Heimel J A F and Coolen A C C 2001 {\em Phys. Rev. E} {\bf 63}
056121
\bibitem{EissOppe92}
Eissfeller H and Opper M 1992 {\em Phys. Rev. Lett.} {\bf 68},
2094
\bibitem{GallCoolSher03} Galla T, Coolen A C C and Sherrington D, {\em J.Phys
A: Math. Gen.} {\bf 36} 11159
\bibitem{HeimDeMa01}
Heimel J A F and De Martino A 2001 {\em J. Phys. A: Math. Gen.}
{\bf 34} L539
\bibitem{DeMa03} De Martino A 2003 {\em Eur. Phys. J. B} {\bf 35} 143 
\bibitem{GallSher05} Galla T and Sherrington D 2005  {\em Eur. Phys. J. B} (at press)
\end{thebibliography}
\end{document}